\documentclass{elsart}

\usepackage[english]{babel}
\usepackage{latexsym}
\usepackage{amsgen,amsbsy,amsopn,amscd,amsmath,amsfonts,amssymb}

\newcommand{\gpf}{\mathcal{E}^{\mathrm{GP}}}
\newcommand{\tff}{\mathcal{E}^{\mathrm{TF}}}
\newcommand{\tfm}{\rho^{\mathrm{TF}}}
\newcommand{\vare}{\varepsilon}
\newcommand{\mb}{\mathbf}
\newcommand{\p}{\partial}
\newcommand{\tb}{\textbf}
\newcommand{\mc}{\mathcal}
\newcommand{\sm}{\setminus}
\newcommand{\tx}{\textit}
\newcommand{\beq}{\begin{equation}}
\newcommand{\eeq}{\end{equation}}
\newcommand{\bdi}{\begin{displaymath}}
\newcommand{\edi}{\end{displaymath}}
\newcommand{\beqn}{\begin{eqnarray}}
\newcommand{\eeqn}{\end{eqnarray}}
\newcommand{\ue}{u_{\varepsilon}}
\newcommand{\fe}{f_{\varepsilon}}
\newcommand{\ve}{v_{\varepsilon}}
\newcommand{\glf}{\mathcal{G}_f}
\newcommand{\rotf}{\mathcal{R}_f}

\begin{document}

\begin{frontmatter}



\title{Vortices in rotating Bose-Einstein condensates confined in homogeneous
traps}

\thanks{Journal Ref.: Physica A 387, 1851-1874 (2008)}


\author{T.Rindler-Daller}

\address{Fakult\"at f\"ur Physik, Universit\"at Wien,
Boltzmanngasse 5, 1090 Vienna, Austria}

\begin{abstract}

We investigate analytically the thermodynamical stability of
vortices in the ground state of rotating 2-dimensional Bose-Einstein
condensates confined in asymptotically homogeneous trapping
potentials in the Thomas-Fermi regime. Our starting point is the
Gross-Pitaevskii energy functional in the rotating frame. By
estimating lower and upper bounds for this energy, we show that the
leading order in energy and density can be described by the
corresponding Thomas-Fermi quantities and we derive the next order
contributions due to vortices. As an application, we consider a
general potential of the form $V(x,y) = (x^2+\lambda^2 y^2)^{s/2}$
with slope $s \in [2,\infty)$ and anisotropy $\lambda \in (0,1]$
which includes the harmonic ($s=2$) and 'flat' ($s \to \infty$)
trap, respectively. For this potential, we derive the critical
angular velocities for the existence of vortices and show that all
vortices are single-quantized. Moreover, we derive relations which
determine the distribution of the vortices in the condensate i.e.
the vortex pattern.

\end{abstract}


\date{}
\end{frontmatter}


\section{Introduction}

Many efforts have been made in understanding ultra-cold quantum
gases, especially since the experimental achievement of
Bose-Einstein condensates (BECs) in 1995. A particular interesting
subject is the study of rotating BECs. When the trap is subjected to
an external rotation the condensate does not rotate like a solid
body. Instead, beyond a critical angular velocity quantized vortices
appear manifesting the genuine quantum character of the system.
Indeed, vortices in BECs were observed in 1999 for the first time
(see Refs. \cite{mat} and \cite{mad,mad2}). Theoretical studies were
already presented before
 (see e.g. \cite{rok} for one of the earliest papers on the subject) and have since then
 grown to a substantial branch of its own
 (see e.g. \cite{aftdan,castin,fetter1,fetter,garcia1,garcia2,garcia3,lund,simula}).
 A general treatment of BECs can be found in the textbooks of
 \cite{PS} and \cite{pita}.\\
 Most of the theoretical studies have been undertaken in the framework of
 the Gross-Pitaevskii (GP) theory whose validity as an approximation of
 the quantum mechanical many-body ground state was established in
 \cite{lieb} for the non-rotating case and in \cite{ls} for rotating
 systems. Particular attention has been put on the so-called Thomas-Fermi (TF) regime of strong
 coupling.
  This is especially true for the study of vortex structures (see the monograph \cite{aftalion}).
   In \cite{ignat1,ignat2} a rigorous
analysis of vortices for
 BECs
  in harmonic anisotropic trap potentials was achieved for a GP-type functional in the TF limit.
   A previous analysis
  was developed in \cite{serfaty} in the context of superfluids. The methodology of those papers originates
   from \cite{BBH}
where a rigorous analysis of vortices in Ginzburg-Landau models of
vanishing magnetic field in the regime which corresponds to the TF
limit was developed.
   In \cite{seir} and
 \cite{seir2}, general results on symmetry breaking which are not limited to the TF regime were proven in
 traps of arbitrary shape.
 \\
 \\
  Within the GP theory, the properties of vortices
  are determined by two physical parameters apart from the external trap, namely angular
velocity and interaction strength between the
 particles.
In this paper, we consider the ground state of rotating 2D
Bose-Einstein condensates which are trapped in asymptotically
homogeneous anisotropic potentials rotating with angular velocity
$\Omega$. The aim of this investigation consists of deducing
analytically the Gross-Pitaevskii energy and density in presence of
vortices and deriving their properties in the TF regime. We consider
thermodynamical conditions for vortex existence, i.e. we are looking
for angular velocities which reduce the total energy in such a way
that vortices are energetically favoured to appear. This work was
originally inspired by the papers of \cite{aftdu} and \cite{castin}
which consider anisotropic harmonic potentials. There and for
instance in Refs. \cite{garcia1,lund,rok}, it was established by
numerical methods that vortices are single-quantized. We show here
by analytical estimates, in particular, that this is true for a very
large class of trapping potentials. In fact, the majority of studies
uses numerical and variational methods for a limited number of trap
potentials (e.g. harmonic or harmonic-plus-quartic) whereas we
derive analytical formulae for a very large class of potentials.
Thereby, we try to present the analysis in such a way that both the
physical ideas and mathematical estimates are brought out in a clear
way.
\\
This paper is organized as follows: In \tx{Section 2}, we state the
setting and present the main result. We decompose the condensate
wave function in a vortex-free part and a vortex-carrying part. This
allows a splitting of the underlying energy functional in separate
contributions which can be estimated subsequently. In \tx{Section
3}, we study the leading asymptotics of the energy and density. In
\tx{Section 4}, we justify a model for the structure and number of
vortex cores which is compatible with the considered order of
magnitude of the angular velocities. \tx{Sections 5-7} contain lower
and upper bound estimates of the vortex-carrying energy
contributions in terms of the winding number of the vortices and the
coupling parameter. In \tx{Section 8}, we specify an external
potential which is of a general anisotropic homogeneous form. For
this potential, we deduce the critical angular velocity for the
appearance of one or a finite number of vortices. The leading orders
of the energy in presence of vortices are calculated and it is shown
that all vortices have winding number one, i.e. they are all
single-quantized. Furthermore, we deduce relations which determine
the distribution of the vortices in the condensate, i.e. the vortex
pattern. Finally, in \tx{Section 9} we present the conclusions.

\section{Setting and main result}

Our starting point is the 2D Gross-Pitaevskii energy functional in
the reference frame rotating (uniformly) with $\tilde{\mb{\Omega}}$
(see e.g. \cite{castin,PS,pita}):
 \beq \label{energie}
  \gpf[u] =  \int_{\mathbb{R}^2} \left[\frac{\hbar^2}{2m}
 |\nabla u|^2 + V(\mb{r})|u|^2 + \frac{Ng}{2}|u|^4 - i u^* \hbar \tilde{\mb{\Omega}} \cdot (\nabla u
 \times \mb{r})\right].
 \eeq
Indeed, it is only meaningful to consider this reference frame as
far as the (temporal) stability of structures is concerned which
appear due to the rotation: The external trap is time-independent
and the states are stationary with respect to that frame (see e.g.
Ref. \cite{castin}).
 The function $u(\mb{r})$ is a complex field (the complex conjugate is denoted as $u^*$).
  We write the associated polar decomposition as $u
  = |u|e^{iS_u}$ where $|u|^2$ is proportional to the density of
  condensed particles with normalization
\bdi
 \int_{\mathbb{R}^2} |u|^2 = 1
 \edi
  and $S_u$ is the phase function. The minimizer of (\ref{energie}) is called the
   order parameter or 'wave function of the
  condensate' in the rotating frame. The external trap potential is denoted by $V$ and $N$ is
  the number of particles with mass $m$. The third term in (\ref{energie}) describes the effective interaction between the particles
  where the coupling constant in
  2D is given by
    $g = \sqrt{8\pi}\hbar^2 a/(m h)$
   with the 3D scattering length $a$ and the thickness $h$
   of the system in the strongly confined direction which we choose to be the $z$-axis,
   so that the system is effectively 2D (in the $x$-$y$-plane).
  We denote $\times$ as
 the vector product in $\mathbb{R}^{3}$, $\mb{r}=(x,y,0)$ and $\tilde{\mb{\Omega}} = (0,0,\tilde{\Omega})$ is the
 angular velocity vector assuming that the gas rotates around the $z$-axis.
 An
important parameter, consisting of the scattering length and a
density, is given by the 'healing length' $\xi$. It is defined
originally by setting
  $\hbar^2/(2m \xi^2) = 2\pi \rho a \hbar^2/m$
  where the r.h.s is the energy per particle for gases in a box
  in the limit of dilute systems $\rho a^3 \to 0$ with density
  $\rho$, so $\xi = 1/\sqrt{4\pi a \rho}$.
    For inhomogeneous and rotating systems, the healing length may be
  defined accordingly by using an appropriate (mean) value for the
  density. In particular, the healing length determines the effective
  radius of a vortex core in rotating systems.
\\
In 2 dimensions, the ratio between the healing length and the
characteristic length of the system $L$, which is set by the
external trap or box respectively, is
 \bdi
  \vare^2 \sim \frac{\xi^2}{L^2} = \frac{\hbar^2}{\sqrt{2\pi}N g
  m}
   \edi
   where we introduce the
dimensionless parameter $\varepsilon$. In this paper, we will be
concerned with the TF limit where this ratio tends to zero (meaning
physically that $0 < \xi \ll L$ or $0 < \vare \ll 1$ respectively).
 However, when
performing the TF limit in a naive way for external potentials,
where the gas can spread out indefinitely, one obtains a trivial
result, namely the minimizer goes to zero and the energy to
infinity. In order to obtain a non-trivial limit, it is then
necessary to rescale all lengths by an $\vare$-dependent factor (see
also \cite{CDY}): Suppose $V$ is homogeneous of order $s$, i.e.
$V(\gamma \mb{r}) = \gamma^s V(\mb{r})$ for $\gamma
> 0$. We rescale the energy functional (\ref{energie}) by setting
$\mb{r} = k\mb{r'}$ and $u(\mb{r}) = u'(\mb{r'})/k$ with $Ng/2 =
\hbar^2/(4\vare^2m)$ and $k= (\hbar^2/(4\vare^2m))^{1/(s+2)}$. Then
we have
  \beq \label{engfunc}
 \gpf[u] = \frac{1}{k^2} \int_{\mathbb{R}^2} \left[ \frac{\hbar^2}{2m}|\nabla' u'|^{2} +
 k^{s+2}V(\mb{r'})|u'|^{2} + \frac{Ng}{2}|u'|^{4} - i\hbar u'^*
 k^2 \tilde{\mb{\Omega}} \cdot (\nabla' u' \times \mb{r'})\right]d^2\mb{r'}
 \eeq
 with $\int |u'|^2 = 1$.
 Choosing $\hbar = 1 = m$ and
inserting $k$,
 (\ref{engfunc}) becomes $\gpf[u] = (16\vare^4)^{1/(s+2)}\mathcal{E}^{\mathrm{GP'}}[u']$
 with the energy on the r.h.s. (omitting the primes)
  \beq \label{skalen}
  \gpf[u] = \int_{\mathbb{R}^2} \left[\frac{1}{2}|\nabla u|^2 +
  \frac{|u|^2}{4\vare^2}(V + |u|^2) - i  u^* \mb{\Omega}(\vare)
  \cdot (\nabla u \times \mb{r})\right]
   \eeq
  and the scaled angular velocity $\mb{\Omega}(\vare)$ is related
  to the original unscaled one by
   \beq \label{skalom}
   \Omega(\vare) = \tilde{\Omega} / (16 \vare^4)^{1/(s+2)}.
   \eeq
  For brevity, we will also write $\Omega$ but it should be
  kept in mind that $\Omega$ depends on $\vare$ after
  scaling. In the forthcoming, we study the functional in (\ref{skalen}) which
  can be also written in the following form
   \beq \label{and}
    \gpf[u] = \int_{\mathbb{R}^2} \left[ \frac{1}{2}|(\nabla -i
    (\mb{\Omega}\times \mb{r}))u|^2 +
    \frac{|u|^2}{4\vare^2}(V+|u|^2) - \frac{1}{2}\Omega^2 r^2
    |u|^2\right]
    \eeq
    and $r := |\mb{r}|$.
 Critical
 points of $\gpf[u]$ are solutions of the following associated Euler-Lagrange
 equation, called Gross-Pitaevskii equation
 \beq
\label{gp}
 \Delta u = \frac{u}{2\vare^2}(V + 2|u|^2 - 4\vare^2\mu^{\rm{GP}}) +
 2i(\mb{\Omega}\times \mb{r})\cdot \nabla u
 \eeq
 where the GP chemical potential $\mu^{\rm{GP}}$ is fixed by the
 normalization. Denoting a minimizer of (\ref{skalen}) as $\ue$,
 it is given by
  \beq \label{mut}
   \mu^{\rm{GP}} = \gpf[\ue] +
   \frac{1}{4\vare^2}\int_{\mathbb{R}^2}|\ue|^4.
    \eeq  The corresponding amplitude squared $|\ue|^2$ will be referred to as Gross-Pitaevskii density.
    Inserting $u = |u|e^{iS_u}$ into (\ref{gp}) results in hydrodynamic-like relations
 for the density and the velocity:
 \bdi
\Delta |u| - |u|(\nabla S_u)^2 + 2|u|(\mb{\Omega}\times \mb{r})\cdot
\nabla S_u - \frac{|u|}{2\vare^2}(V + 2 |u|^2 -
4\vare^2\mu^{\rm{GP}}) = 0,
 \edi
  \bdi
\nabla \cdot [ |u|^2(\nabla S_u - \mb{\Omega}\times \mb{r})] = 0.
 \edi
  The GP functional (\ref{skalen}) for $\Omega = 0$ decribes the gas without rotation
\beq \label{gpohn}
 \gpf[f] = \int_{\mathbb{R}^2} \left[\frac{1}{2}(\nabla f)^2 +
 \frac{f^2}{4\vare^2}(V+f^2)\right]
 \eeq
with $f$ a real, positive function. The minimizer of (\ref{gpohn})
will be denoted as $\fe$. The normalization condition
$\int_{\mathbb{R}^2} f^2 = 1$ fixes the associated chemical
 potential $\nu^{\rm{GP}}$ which is given by
  \beq \label{nuc}
   \nu^{\rm{GP}} = \gpf[\fe] + \frac{1}{4\vare^2}\int \fe^4
   \eeq
  and which is of the order $1/\vare^2$. The functional
  (\ref{gpohn})
   tends for $\vare \to 0$ to
a Thomas-Fermi type functional
 \beq \label{tffunc}
 \tff[\rho] = \frac{1}{4\vare^2}\int_{\mathbb{R}^2}\rho (V+\rho),
 \eeq
 which is a functional for the density
$\rho=f^2$ alone.
 It can be shown (see e.g. Ref. \cite{lieb}) that it has a unique positive
 minimizer,
 the Thomas-Fermi density,
 \beq \label{tfmin}
  \tfm = \frac{1}{2}[4\vare^2\mu^{\rm{TF}} - V]_+ =: \frac{1}{2}[\mu -
  V]_+
  \eeq
   where $[.]_+$ denotes the positive part and $\mu := 4\vare^2 \mu^{\rm{TF}}$.
  The TF chemical potential $\mu^{\rm{TF}}$ (or $\mu$ respectively) is determined by
\beq \label{norm}
   \int_{\mc{D}} \tfm = 1
    \eeq
  where
  \bdi
   \mc{D} = \{(x,y) \in \mathbb{R}^2: \tfm > 0 \}
    \edi
   is the Thomas-Fermi domain whose shape depends on the external potential $V$.
Moreover,
   \bdi
  \mu^{\rm{TF}} = \tff[\tfm] + \frac{1}{4\vare^2}\int
  (\tfm)^2,
   \edi
  which is of the order $1/\vare^2$ whereas $\mu := 4\vare^2 \mu^{\rm{TF}}$
 is of the order of a constant independent of $\vare$.

\subsection{Splitting of the GP energy functional}

 In the TF regime where $\vare$
 is small, vortex cores are small compared to the characteristic length scale of the system,
 producing narrow 'holes' which effectively shrink as $\vare \to 0$. It is argued in \tx{Section
4} that vortices appear at a critical angular velocity of the order
$\Omega \simeq C |\ln \vare|$ with $C$ a positive constant
(independent of $\vare$) depending on the external trap. Explicit
expressions for $C$ will be determined in the forthcoming analysis
(see also Refs. \cite{aftdu} and \cite{castin} for the harmonic
trap case). \\
In the minimization of (\ref{gpohn}), i.e. (\ref{skalen}) with
$\Omega = 0$, one considers all functions in the subspace of angular
momentum zero and the density profile is given by $\fe^2$.
Considering (\ref{skalen}) with $\Omega > 0$ we will see that, as
long as $\Omega \leq C |\ln \vare|$ asymptotically, the overall
density can still be described by the vortex-free density $\fe^2$ in
good approximation. However, in a non-isotropic potential $V$ there
appears a phase $S$ (depending on $V$), i.e. the vortex-free
function is then more generally $fe^{iS}$. Since this function has
no vortex, the phase $S$ is non-singular and (\ref{gp}) gives \beq
\label{fsa} \Delta f = - f \nabla S \cdot [2(\mb{\Omega} \times
\mb{r}) - \nabla S] + \frac{f}{2\vare^2}(V+2f^2-4\vare^2
\tilde{\nu}^{\rm{GP}}) \eeq and \beq \label{conta} \nabla \cdot
[f^2(\nabla S -\mb{\Omega} \times \mb{r})] = 0 \eeq
 where $\tilde{\nu}^{\rm{GP}}$ is the associated chemical potential.
 A solution without vortex is a minimizer of the problem $\min
\{\gpf[fe^{iS}]: fe^{iS} \in H^1 \mbox{ with } f>0, \int f^2 = 1\}$
 (see also \cite{ignat1}) with
\beq \label{vofree} \gpf[fe^{iS}] = \int_{\mathbb{R}^2} \left[
\frac{1}{2}(\nabla f)^2 + \frac{f^2}{4\vare^2}(V+f^2) +
\frac{1}{2}f^2[(\nabla S)^2 - 2 \nabla S \cdot (\mb{\Omega}\times
\mb{r})]  \right].
 \eeq
Later on in this paper we are going to consider external traps of
the form
 \beq \label{potential}
 V(x,y) = (x^2+\lambda^2 y^2)^{s/2}
 \eeq
 with slope $s \in [2,\infty)$ and $\lambda \in (0,1]$ describing the anisotropy.
 It is a fairly general potential which includes also the
important special cases of the harmonic ($s=2$) and flat ($s \to
\infty$) trap which are extensively used in experiments. The
corresponding phase to
 this potential is
  \beq \label{nosing}
  S =\frac{\lambda^2-1}{\lambda^2+1}\Omega x y
  \eeq
which vanishes for the isotropic case $\lambda = 1$. This expression
for $S$ was also deduced for the harmonic trap in Refs. \cite{aftdu}
and \cite{castin}. Note, however, that it is not dependent on the
slope parameter $s$. We also see that the terms in (\ref{vofree})
involving $\nabla S$ are at most of the order $|\ln \vare|^2$ for
$\Omega \leq |\ln \vare|$ and hence of much lower order than the
remaining part described by (\ref{gpohn}) which is $\sim 1/\vare^2$.
 \\
We now decompose the order parameter $u$ of (\ref{skalen}) into the
vortex-free part $fe^{iS}$ and a part which carries the vorticity. A
similar splitting can be found in Refs.
\cite{aftdu,andre,castin,kav} and more recently in Refs.
\cite{baym,FB,WGBP}. Writing
 $u = |u|e^{iS_u} = fe^{iS} v =
f|v|e^{i(S+S_v)}$
  with $|u|=f|v|$ and $S_u=S+S_v$, the contribution
$v = |v|e^{iS_v}$ accounts for the presence of vortices. In a vortex
point, the amplitude vanishes, i.e. $|u|= |v| = 0$ since $f \not= 0$
and the phase fulfills the usual circulation condition which is a
quantization condition because $u$ (resp. $v$) is a complex field:
 \bdi
\oint_{\mc{C}} \nabla S_u \cdot \tau = \oint_{\mc{C}} (\nabla S_v +
\nabla S) \cdot \tau = 2\pi d + 0 \edi since $S$ has no singularity
and $\tau$ is a unit tangent vector to the curve $\mc{C}$ encircling
the vortex with winding number $d$. Without the presence of
vortices, there would be
 $u = fe^{iS}$ with density $|u|^2=f^2$ and the phase $S_u$ would be non-singular.
Inserting the decomposition $u = fe^{iS}v$ in the energy functional
(\ref{skalen}) results in the following splitting (see also
\cite{aftdu,andre}) where the integrals are over $\mathbb{R}^2$: The
first term becomes \bdi \int \frac{1}{2}|\nabla (fe^{iS} v)|^2 =
\int \left[\frac{1}{2}f^2|\nabla v|^2 + \frac{1}{2}|v|^2[(\nabla
f)^2+f^2(\nabla S)^2] +\frac{1}{4}\nabla(f^2) \cdot \nabla |v|^2
\right.\edi \bdi
 \left.+ \frac{1}{2}f^2\nabla S \cdot (iv\nabla v^* -
i v^* \nabla v)\right], \edi the second one is simply
 \bdi
 \int \frac{|fe^{iS}v|^2}{4\vare^2}(V + |fe^{iS}v|^2) =
 \int \frac{f^2|v|^2}{4\vare^2}(V+f^2|v|^2)
   \edi
and for the rotation term we get \bdi  -\int i fe^{-iS} v^*
\mb{\Omega} \cdot (\nabla (fe^{iS} v) \times \mb{r}) = \int i f^2
v^* \nabla v \cdot (\mb{\Omega} \times \mb{r}) - \int f^2|v|^2\nabla
S \cdot (\mb{\Omega}\times \mb{r}).
 \edi
Putting the terms together and separating the vortex-free part of
the energy (\ref{vofree}), we have
 \beqn
\label{eueta} \lefteqn{\gpf[u] = \gpf[fe^{iS}] + \int (|v|^2-1)
\times {}}
\nonumber\\
& &{} \left[\frac{1}{2}(\nabla f)^2 + \frac{1}{2}f^2(\nabla S)^2 -
f^2\nabla S \cdot (\mb{\Omega}\times \mb{r}) +
\frac{Vf^2}{4\vare^2}\right] +
 \nonumber\\
& &{} + \int \frac{1}{4}\nabla(f^2) \cdot \nabla |v|^2 + \int
\frac{1}{2}f^2|\nabla v|^2 + \int
\frac{f^4}{4\vare^2}(|v|^4-1) + \nonumber\\
& &{} +  \int \left[\frac{1}{2}f^2\nabla S (i v \nabla  v^* - i v^*
\nabla v) + i f^2 v^* \nabla v \cdot (\mb{\Omega}\times \mb{r})
\right]. \eeqn The third term of this expression becomes
 \bdi  \int
\frac{1}{4}\nabla(f^2) \cdot \nabla |v|^2 = \int
\frac{1}{4}\nabla(f^2) \cdot \nabla (|v|^2-1) = -\int
\frac{1}{4}(|v|^2-1)\Delta(f^2) \edi
 \bdi  = -\int
\frac{1}{2}(|v|^2-1)f\Delta f - \int \frac{1}{2}(|v|^2-1)(\nabla
f)^2 \edi \bdi =  \int (|v|^2-1)\left[f^2\nabla S \cdot
(\mb{\Omega}\times \mb{r}) - \frac{1}{2}f^2(\nabla S)^2 -
\frac{f^2}{4\vare^2}(V + 2f^2 - 4\vare^2\tilde{\nu}^{\rm{GP}}) -
\frac{1}{2}(\nabla f)^2 \right], \edi
where we used (\ref{fsa}) for $\Delta f$.\\
Moreover, for the fifth term in (\ref{eueta}) we use the identity
\bdi \int \frac{f^4}{4\vare^2}(|v|^4-1) = \int
\frac{f^4}{2\vare^2}(|v|^2-1) + \int
\frac{f^4}{4\vare^2}(1-|v|^2)^2. \edi Inserting the last two
equations into (\ref{eueta}) we get the following splitting of the
functional in (\ref{skalen})
 \bdi
\gpf[u] = \gpf[fe^{iS}] + \int_{\mathbb{R}^2}
\left[\frac{f^2}{2}|\nabla v|^2 +
\frac{f^4}{4\vare^2}(1-|v|^2)^2\right] - \int_{\mathbb{R}^2} if^2
v^* \nabla v \cdot (\nabla S - \mb{\Omega}\times \mb{r}) \edi \beq
\label{splitting} =: \gpf[fe^{iS}] + \mc{G}_f[v] - \mc{R}_f[v] \eeq
where we used $\tilde{\nu}^{\rm{GP}} \int f^2(|v|^2-1) = 0$ because
of the normalization conditions and the last term in (\ref{eueta})
was written in a more convenient form using \bdi \int
\left[\frac{1}{2}f^2\nabla S \cdot (iv\nabla  v^* - i v^* \nabla v)
+ f^2i v^* \nabla v \cdot (\mb{\Omega}\times \mb{r})\right]
 \edi
\bdi = \int f^2 \nabla S \cdot (iv,\nabla v) - \int \left[f^2
(\mb{\Omega}\times \mb{r})\cdot (iv,\nabla v) +
\frac{f^2}{2}i(\mb{\Omega}\times \mb{r})\cdot \nabla (|v|^2)\right]
\edi
 \bdi = \int f^2(iv,\nabla v)\cdot (\nabla S - \mb{\Omega}\times
\mb{r}) = -\int f^2i v^* \nabla v \cdot (\nabla S -
\mb{\Omega}\times \mb{r}) \edi where $(u,v) := (u v^* + u^* v)/2$.
 The terms apart from the vortex-free energy in (\ref{splitting}) describe the contribution of the vorticity to the energy:
  The second term $\mc{G}_f[v]$ looks
formally like a 'weighted' Ginzburg-Landau (GL) energy functional
without magnetic field and accordingly will be called GL-type energy
in the forthcoming and
 $\mc{R}_f[v]$ is the rotation energy.\\
Using the splitting (\ref{splitting}), vortices of $u$ (if present)
are vortices of $v$ and they are described via the functionals
$\mc{G}_f[v] - \mc{R}_f[v]$.

\subsection{Main result}

We have the following main result:
\\
\\
\tb{Main result:} Let $\ue$ be a minimizer of (\ref{skalen}) and
$\fe$ a minimizer of (\ref{vofree}) for $V$ in (\ref{potential}) and
$S$ in (\ref{nosing}) and under the normalization constraints. Let
$C$ and $\delta$ be positive constants independent
of $\vare$ with $0< \delta \ll 1$ and let $o(1)$ denote a quantity which goes to zero as $\vare \to 0$.\\
For some integer $n\geq 1$ and
 \beq \label{herz}
 \Omega_n = C_1 \left[|\ln \vare| + (n-1) \ln |\ln
 \vare|\right] =: \Omega_1 + C_1 (n-1) \ln |\ln \vare|
  \eeq
 with
 \bdi
 C_1 := \frac{s+2}{s\mu^{2/s}}~\frac{1+\lambda^2}{2},
 \edi
we have the following results: \\
 i) If $\Omega \leq \Omega_1 - C_1\delta \ln
|\ln \vare|$ and $\vare$ sufficiently small, then $\ue$ has no
vortices in $\mc{D} \sm \p \mc{D}$ and the Gross-Pitaevskii energy
is
 \beq \label{enga}
  \gpf[\ue] = \gpf[\fe e^{iS}] + C.
 \eeq
ii) If $\Omega_n + C_1\delta \ln |\ln \vare| \leq \Omega \leq
\Omega_{n+1} - C_1\delta \ln |\ln \vare|$ for $n\geq 1$,
   then, for $\vare$ sufficiently small, $\ue$ has $n$ vortices with
   winding number
one located in $\mb{r}_1,...,\mb{r}_n \in \mc{D} \sm \p \mc{D}$,
$\mb{r}_i=(x_i,y_i), i=1,..,n$. Setting $\tilde{\mb{r}}_i =
(\tilde{x}_i,\tilde{y}_i)$ with $\tilde{x}_i = x_i \sqrt{\Omega},
\tilde{y}_i=y_i \lambda \sqrt{\Omega}$, the configuration
($\tilde{\mb{r}}_1,...,\tilde{\mb{r}}_n$) minimizes the function
 \bdi
  w(\mb{a}_1,..,\mb{a}_n) = -\frac{\pi \mu}{4}\sum_{i \not=
  j} \ln [(X_i - X_j)^2 +
  \lambda^{-2}(Y_i-Y_j)^2] + \frac{\pi
  \mu}{1+\lambda^2} \sum_{i=1}^n (X_i^2+Y_i^2) -
   \edi
   \bdi -
  \frac{\pi \ln \Omega}{4\Omega^{s/2}} \sum_{i=1}^n
  (X_i^2+Y_i^2)^{s/2}
  \edi
  with $\mb{a}_i = (X_i, Y_i), i=1,..,n$
and the Gross-Pitaevskii energy is
 \bdi
 \gpf[\ue] = \gpf[\fe e^{iS}] + \frac{\pi}{2}\mu n \left(|\ln \vare| -
 \frac{2s}{(1+\lambda^2)(s+2)}\mu^{2/s}\Omega\right) +
  \edi
   \beq \label{nenergie} +
 \frac{\pi}{4}\mu n (n-1) \ln \Omega + w(\tilde{\mb{r}}_1,..,\tilde{\mb{r}}_n) + C + o(1).
 \eeq
The proof is split into several estimates which are shown in the
following sections. There, positive constants are denoted by $C$
(sometimes carrying primes) and they may change from line to line.

\section{The leading order in energy and density}

In this section, we show the leading asymptotics for the GP energy
and density. We will see, in particular, that it is not affected by
vortices whose influence can only be seen in the next lower order.
The leading term in the energy comes from the TF contribution in
(\ref{tffunc}) which is $\sim 1/\vare^2$ whereas vortices contribute
to the order $\Omega \sim |\ln \vare|$ (see also \tx{Section 4}).
However, the determination of the precise expressions in
(\ref{nenergie}) requires a more detailed analysis
which is carried out in \tx{Sections 5-8}. \\
For the following estimates, we introduce the function
 \beq \label{afunk}
  b(\mb{r}) := \frac{1}{2}(\mu - V(\mb{r}))
   \eeq
 whose positive part is the TF density, i.e. $[b(\mb{r})]_+ := \tfm$.
\\
\\
\tb{Estimate 1:}
 For $\Omega(\vare)$ satisfying $C_V |\ln \vare| \leq \Omega(\vare) < C |\ln \vare|$
 where $C_V$ depends on the parameters of the external potential
    $V$, $C > C_V$,
    and for \( \varepsilon \) sufficiently small,
        \beq
            \label{energynorothom}
            \gpf[\ue] = \tff[\tfm] + C|\ln \vare|
        \eeq
    and
        \bdi
            \int_{\mathbb{R}^2} \left(|\ue|^2 - \tfm \right)^2 = o(1).
        \edi
\tx{Proof:}\\
 This can be shown similar as Prop. 2.3 in \cite{CDY}. The
lower bound can be trivially obtained by neglecting the first
positive term in (\ref{and})
 \bdi
   \gpf[\ue] \geq \tff[\tfm] - C \Omega(\vare)^2.
   \edi
The upper bound can be obtained by using $\gpf[\ue] \leq
\gpf[\ue]|_{\Omega=0}$ and $\sqrt{\tfm}$ as a trial function,
 \bdi
  \gpf[\ue]|_{\Omega=0} \leq \tff[\tfm] + C |\ln \vare|.
   \edi
Concerning the density asymptotics, we estimate the following: using
the negativity of $b(\mb{r})$ outside the TF domain, we have
 \bdi
 \int_{\mathbb{R}^2} (|\ue|^2 - \tfm)^2 \leq \int_{\mathbb{R}^2} \left(|\ue|^4 -
 2 b(\mb{r}) |\ue|^2 + (\tfm)^2 \right).
 \edi
 On the other hand, we deduce
 \bdi
  4\vare^2\tff[|\ue|^2] = \int_{\mathbb{R}^2} \left[|\ue|^4 +
  |\ue|^2 V \right]
   =
    \int_{\mathbb{R}^2} |\ue|^4 -2
  \int_{\mathbb{R}^2} b(\mb{r}) |\ue|^2  + \mu,
\edi that is \bdi
 \int_{\mathbb{R}^2} (|\ue|^2 - \tfm)^2 \leq 4\vare^2\tff[|\ue|^2] +
 \int_{\mathbb{R}^2} (\tfm)^2 - \mu
 \edi
  \bdi
   = 4\vare^2 (\tff[|\ue|^2] - \tff[\tfm])  \leq C \vare^2 |\ln
   \vare|
   \edi
   and the last inequality follows from
   (\ref{energynorothom}). Thus, the GP density approaches the TF
   density for $\vare \to 0$ showing \tx{Estimate 1}. (A similar result is also true for higher angular
   velocities as is shown in \cite{CDY}).
In the same way, we see the following result which is used to show
\tx{Estimate 3} below
 \bdi
        \int_{\mc{D}} \left( |\ue|^2 - \tfm \right)^2 + \int_{\mathbb{R}^2 \setminus \mc{D}}
         | \ue |^4 = 4\vare^2 \tff[|\ue|^2] + \int_{\mc{D}}(\tfm)^2
        - \mu
    \edi
    \bdi
        \leq 4\vare^2 (\tff[|\ue|^2] - \tff[\tfm]) \leq  C\varepsilon^2 |\ln \vare|,
    \edi
    so that
    \beq
        \label{l4normhom}
            \int_{\mathbb{R}^2 \setminus \mc{D}} | \ue |^4 \leq C \varepsilon^2 |\ln \vare|.
        \eeq
        \\
        \\
 Now we return to
the non-rotating ground state described by (\ref{gpohn}).  We have
the following point-wise estimate for $\fe$ within the TF domain:
\\
\\
\tb{Estimate 2:}
        Let \( f_{\varepsilon} \) be a minimizer of (\ref{gpohn}) under the normalization constraint.
        It is the unique
        positive solution of
  \beq \label{eulf}
 \Delta f = \frac{f}{2\vare^{2}}(V+2f^{2}-4\vare^2 \nu^{\rm{GP}})
 ~~~\rm{in}~~ \mathbb{R}^2
 \eeq
 with the chemical potential $\nu^{\rm{GP}}$ in (\ref{nuc}).
 If
$\vare$ is sufficiently small, then
 \beq \label{pointwisa}
  |\sqrt{\tfm(\mb{r})} - \fe (\mb{r})| \leq C \vare^{1/3} \sqrt{\tfm
  (\mb{r})}
  \eeq
 for $\mb{r} \in \mc{D}^{\rm{in}}:=\{\mb{r} \in \mathbb{R}^2: V(\mb{r}) \leq \mu - \vare^{1/3}\}$.
 That is, we may replace the vortex-free density $\fe^2$ by the
 Thomas-Fermi
 density $\tfm$ within a region almost as large as the Thomas-Fermi domain making only an error of order $o(1)$.
\\
\\
\tx{Proof:}\\
 As is shown in Ref. \cite{lieb}, there exists a unique
minimizer for the functional (\ref{gpohn}).
 Since each minimizer fulfills
 (\ref{eulf}) (which is the corresponding Euler-Lagrange equation)
 and $\gpf[f_{\vare}] = \gpf[|f_{\vare}|]$ the
 positivity of the minimizer $f_{\vare}$ follows.
 Now  we  look at (\ref{pointwisa}). It can be shown similar as in Refs. \cite{afta,andre} by using
 suitable sub- and supersolutions:
We consider a disc $B_{\delta}(\mb{r}_0)$ around $\mb{r}_0 \in
\mc{D}':=\{\mb{r} \in \mathbb{R}^2: V(\mb{r}) \leq \mu - t, t>0\}$
with radius $\delta < t$ and construct a subsolution $w(\mb{r}) =
\sqrt{\rho} \tanh q$ with $\mb{r} \in B_{\delta}(\mb{r}_0)$, $q :=
\frac{\delta^2-|\mb{r}-\mb{r}_0|^2}{\delta \vare}$ and $\rho :=
\min_{B_{\delta(\mb{r}_0)}} \tfm$.
 Using $w(\mb{r})$, we see that
  $\Delta w \geq
\frac{w}{2\vare^2}\left(V + 2w^2 -4\vare^2 \nu^{\rm{GP}}\right)$
 is fulfilled since
   $4\vare^2 \nu^{\rm{GP}} > \mu + 2[\rho \tanh^2 q - \tfm]$
  for $\vare$ sufficiently small.
 On $\p
B_{\delta}(\mb{r}_0)$ there is $|\mb{r}-\mb{r}_0|=\delta$ and
$w|_{\p B_{\delta}} = 0 < \fe$. So $w$ is a subsolution for
(\ref{eulf}) in $B_{\delta}(\mb{r}_0)$ and \beq \sqrt{\rho} -
\fe(\mb{r}_0) \leq \sqrt{\rho} - w(\mb{r}_0) =
\frac{2\sqrt{\rho}e^{-2\delta/\vare}}{1+e^{-2\delta/\vare}} \leq
 2\sqrt{\rho}e^{-2\delta/\vare} \leq 2\sqrt{\tfm}e^{-2\delta/\vare}.
\eeq
 Since $\tfm$ is smooth in $\mc{D} \sm \p \mc{D}$, we can approximate
$\tfm(\mb{r}_0)$ by $\rho$, making a small error of the order
$o(1)$. So, \bdi
\frac{\sqrt{\tfm(\mb{r}_0)}-\fe(\mb{r}_0)}{\sqrt{\tfm(\mb{r}_0)}}
\leq \frac{\sqrt{\tfm(\mb{r}_0)}-\sqrt{\rho}}{\sqrt{\tfm(\mb{r}_0)}}
 + \frac{2\sqrt{\rho}}{\sqrt{\tfm(\mb{r}_0)}}e^{-2\delta/\vare}
 \leq C\left(\frac{\delta}{\sqrt{t}} + e^{-2\delta/\vare}\right).
\edi $\delta$ must be chosen such that $e^{-2\delta/\vare}$ is
exponentially small. We choose $\delta = \vare^{2/3}$ and $t =
\vare^{1/3}$ as in \cite{afta}. Likewise we construct a
supersolution
 $p(\mb{r}) = \sqrt{m}\coth [\rm{arcoth} (\sqrt{M/m})
 + \sqrt{m} q]$
  with $\mb{r} \in B_{\delta}(\mb{r}_0)$, $m := \max_{B_{\delta(\mb{r}_0)}} \tfm$,
  $M = \max_{\mc{D}} \tfm$ and $q$ as above.
Using $p(\mb{r})$, we see that
   $\Delta p \leq
\frac{p}{2\vare^2}\left(V + 2p^2 -4\vare^2 \nu^{\rm{GP}}\right)$
  is fulfilled since
    $4\vare^2 \nu^{\rm{GP}} \leq \mu + 2(m \coth^2
    (\rm{arcoth} (\sqrt{M/m}) + \sqrt{m}q) -
    \tfm)$
 for $\vare$ sufficiently small. On $\p B_{\delta}(\mb{r}_0)$ there
is $|\mb{r}-\mb{r}_0| = \delta$ and $p|_{\p B_{\delta}} = \sqrt{M}
\geq \fe$. So $p(\mb{r})$ is a
 supersolution for (\ref{eulf}) in $B_{\delta}(\mb{r}_0)$. Proceeding
 as before, we get \bdi
\frac{\fe(\mb{r}_0)-\sqrt{\tfm(\mb{r}_0)}}{\sqrt{\tfm(\mb{r}_0)}}
\leq \frac{p(\mb{r}_0)-\sqrt{\tfm(\mb{r}_0)}}{\sqrt{\tfm(\mb{r}_0)}}
\leq C\left(\frac{\delta}{\sqrt{t}} + e^{-2\delta/\vare}\right).
\edi Choosing $\delta$ and $t$ appropriately again, we get
(\ref{pointwisa}) for any $\mb{r} \in \mc{D}^{\rm{in}}$.
\\
\\
It is intuitively clear that only the vortex-free density $\fe^2$
and not the 'full' GP density $|\ue|^2$ can satisfy a pointwise
estimate as above: $\ue$ may have vortices whereas $\tfm$ carries no
vorticity at all. However, what can be shown is the fact that, in
the TF regime, where $\vare \to 0$, $|\ue|^2$ is exponentially small
outside the TF domain (see also Refs. \cite{ignat1} and \cite{CDY}):
\\
\\
\tb{Estimate 3:} For $\mb{r} \in \Theta_{\vare}:=\{\mb{r} \in
\mathbb{R}^2: V(\mb{r}) > \mu + \vare^{1/3}\}$ and $\vare$
sufficiently small, there is
 \bdi
  |\ue(\mb{r})|^2 \leq C \vare^{1/6}|\ln \vare|^{1/2}\exp
  \left(\frac {b(\mb{r})}{C\vare^{2/3}}\right),
   \edi
  where $b(\mb{r})$ is defined in (\ref{afunk}).
  \\
  \\
  \tx{Proof:} \\
  By using (\ref{gp}) we have
   \bdi
    -\frac{1}{2}\Delta |\ue|^2 = -|\nabla \ue|^2 -
    \frac{V}{2\vare^2}|\ue|^2 - \frac{|\ue|^4}{\vare^2} +
    2\mu^{\rm{GP}}|\ue|^2 - i(\ue (\mb{\Omega}\times \mb{r})\cdot \nabla \ue^* + \ue^*
    (\mb{\Omega}\times \mb{r})\cdot \nabla \ue).
     \edi
   The estimate
    \bdi
      2\Omega(\vare) |i\ue^* \nabla \ue \times \mb{r}| \leq |\nabla \ue|^2 +
      \Omega(\vare)^2|\mb{r}|^2|\ue|^2
      \edi
   leads to
\bdi
            -\frac{1}{2} \Delta |\ue|^2 \leq \left[ \Omega(\vare)^2 |\mb{r}|^2 \vare^2
            - \frac{V}{2} - |\ue|^2 + 2\mu^{\rm{GP}}\vare^2\right] \frac{|\ue|^2}{\varepsilon^2}
            \mbox{ in } \mathbb{R}^2.
        \edi
From (\ref{mut}) and \tx{Estimate 1} follows
 \bdi
  \vare^2
\mu^{\rm{GP}} = \vare^2 \gpf[\ue] +
\frac{1}{4}\int_{\mathbb{R}^2}|\ue|^4 \leq \vare^2 \tff[\tfm] + o(1)
+ \frac{1}{4}\int_{\mathbb{R}^2}|\ue|^4
 \edi
 \bdi
  = \vare^2\mu^{\rm{TF}} + o(1) + \frac{1}{4}\int_{\mathbb{R}^2}(|\ue|^4 - (\tfm)^2)
  \leq \vare^2 \mu^{\rm{TF}} + o(1). \edi
  Thus
 \bdi
-\frac{1}{2} \Delta |\ue|^2 \leq \left[ \Omega(\vare)^2
|\mb{r}|^2\vare^2 - \frac{V}{2} + 2\vare^2\mu^{\rm{TF}} + o(1) -
|\ue|^2 \right] \frac{|\ue|^2}{\varepsilon^2}
  \leq C \frac{b(\mb{r})}{\vare^2}|\ue|^2 < 0
\edi
 in $\Theta_{\vare}$ where $b(\mb{r}) <
 -\frac{\vare^{1/3}}{4}$. That is
$|\ue|^2$ fulfills
 \beq \label{unu}
  -\vare^2 \Delta |\ue|^2 - C'b(\mb{r})|\ue|^2 \leq 0 \mbox{ in
  } \Theta_{\vare}.
   \eeq
So $|\ue|^2$ is subharmonic in $\Theta_{\vare}$ for $\vare$
sufficiently small. That means, there is for all $r=|\mb{r}|$ with
$B_{\varrho}(r) \subset \Theta_{\vare}$ that
 \bdi
  |\ue (\mb{r})|^2
\leq \frac{1}{\pi \varrho^2}\int_{B_{\varrho}(r)}|\ue|^2 \leq
\frac{1}{\sqrt{\pi} \varrho}\left(\int_{\mb{r}\in
\Theta_{\vare}}|\ue|^4\right)^{1/2} \leq
\frac{C}{\varrho}\vare^{1/2}|\ln \vare|^{1/2} \edi using
(\ref{l4normhom}). If we now take $\mb{r} \in \Sigma_{\vare} :=
\{\mb{r} \in \mathbb{R}^2: V(\mb{r}) \geq \mu +
\frac{\vare^{1/3}}{2}\}$ and choose $\varrho =
\frac{\vare^{1/3}}{2}$ we get \bdi |\ue (\mb{r})|^2 \leq
C\vare^{1/6}|\ln \vare|^{1/2} \edi so that $|\ue (\mb{r})|^2 \to 0$
in $\Sigma_{\vare}$ for $\vare \to 0$. Moreover, from (\ref{unu}) it
follows that $|\ue|^2$ is a subsolution of \beq \label{her}
            \left\{
            \begin{array}{l}
                    - \Delta w + C''\vare^{-5/3}w = 0 \mbox{ in } \Sigma_{\vare}   \\
                    \mbox{} \\
                    w = C\vare^{1/6}|\ln \vare|^{1/2} \mbox{ on } \p \Sigma_{\vare}.
            \end{array}
            \right.
        \eeq
On the other hand, one can verify that \bdi \tilde{u} =
C\vare^{1/6}|\ln
\vare|^{1/2}\exp\left(\frac{b(\mb{r})}{C\vare^{2/3}}\right) \edi is
a supersolution of (\ref{her}). Therefore $0 \leq |\ue (\mb{r})|^2
\leq \tilde{u}$ for $\mb{r} \in \Theta_{\vare}$.
\\
\\
 So, since $|\ue|^2$ is exponentially small in $\vare$ outside of the TF domain $\mc{D}$, the
 above
energy splitting (\ref{splitting}) can be put now into the form \beq
\label{tfsplitting} \gpf[\ue] = \gpf[\fe e^{iS}] + \int_{\mc{D}}
\left[\frac{\fe^2}{2}|\nabla \ve|^2 +
\frac{\fe^4}{4\vare^2}(1-|\ve|^2)^2\right] - \int_{\mc{D}} i\fe^2
\ve^* \nabla \ve \cdot (\nabla S - \mb{\Omega}\times \mb{r}) + o(1)
\eeq \bdi =: \gpf[\fe e^{iS}] + \glf[\ve] - \rotf[\ve] + o(1) \edi
where $\ve = \ue / \fe e^{iS}$ and $o(1) \to 0$ as $\vare \to 0$.
Thus in the following, it suffices to restrict our considerations to
the Thomas-Fermi domain $\mc{D}$.

\section{Vorticity in the Thomas-Fermi regime}

We know from experiments that angular momentum is quantized in the
form of vortices when the gas is subjected to an external rotation.
Hence we may approximate the vorticity field by $N_v$ isolated point
vortices. However, it is a difficult task in general to prove the
validity of this approximation rigorously from more basic
properties. It has been shown in the work of \cite{ignat2} for BECs
in harmonic anisotropic traps that the vorticity is indeed
concentrated in a finite (independent of $\vare$) number of vortex
cores if one assumes that the angular velocity is bounded by $\Omega
\leq C |\ln \vare|$ asymptotically. This has been achieved by using
a number of technical vortex core constructions. We will not
generalize these methods to the more general traps considered here
but instead we like to argue by physical reasoning how the number of
vortices scales with $\Omega(\vare)$.
\\
In experiments $\Omega$ and $\vare$ are independent parameters.
Usually, the interaction between the particles is tuned and
afterwards $\Omega$ is increased (independently of $\vare$) beyond
the critical value. So in principle one could study the whole
parameter domain spanned by $\Omega$ and (here) positive $\vare$.
However, we restricted to the TF regime where the scaled angular
velocity $\Omega = \Omega(\vare)$ depends on $\vare$ in such a way
that for $\vare \to 0$, $\Omega \to \infty$ and hence we cover only
a fraction of the possible parameter domain.
 Which dependence of $\Omega(\vare)$ may occur in the TF regime ?  There are
 essentially three regimes in $\Omega$ for non-harmonic
traps
 where interesting effects appear (see \cite{CDY}), namely
  $\Omega \sim |\ln \vare|, \Omega \sim 1/\vare, \Omega \gg 1/\vare$ (the first regime also applies to
  harmonic traps).
  One may ask for a connection between different vortex core sizes,
  the magnitude of $\Omega(\vare)$ and the kind of defects appearing
in the condensate.
   For $\Omega \sim |\ln \vare|$, one may deduce similar estimates
   for the vortex energy
   using core sizes of the order
   $\sigma = \kappa \vare$ or $\sigma = \vare^{\alpha}$ with constants $\kappa, \alpha > 0$ and the choice is
   fixed by technical reasons.
   However, in the fast rotating regimes, the size of the defects
   seems
to be much more restrictive. As is shown in \cite{CDY}, for $\Omega
\sim 1/\vare$ there appears a 'hole' around the origin and the core
size of the vortices itself is of the order $\sqrt{\vare}$. For even
larger velocities $\Omega \gg 1/\vare$, the condensate is expelled
to a small layer at the boundary and there remains a 'giant vortex'
state filling out almost all of the condensate.
 \footnote{One may argue that vortices with larger core
radii, say e.g. $\sigma \sim 1/|\ln \vare|$ could in principle exist
at lower angular velocities of the order $\Omega \sim \ln |\ln
\vare|$. However, the characteristic length, where perturbations of
the condensate wave function are smoothed out, is given by the
healing length $\xi$ or $\vare$ respectively. Hence we expect the
cores to be of the order $\vare$ and larger cores are not stable in
the setting described here. Moreover, for angular velocities of the
order $|\ln \vare|$ we may also not expect the appearance of
pathological cases like non-isolated vortices forming dense
1-dimensional structures because they would have a much higher
energy than would be favourable at this order of $\Omega$. It seems
that the underlying equations are too regular to support such kinds
of defects even at much higher angular velocities.}
\\
 In the regime of large vorticity one may also
consider a kind of correspondence principle for a large number of
vortices which is argued by Feynman \cite{feyn} in the context of
rotating superfluid $^4$He: a dense lattice of uniform distributed
vortices should 'mimic' solid-body rotation on average, although the
flow is strictly irrotational away from the vortex cores. The
circulation around a closed contour $\mc{C}$ which encloses a large
number of vortices $N_v$ is $\Gamma = \oint_{\mc{C}} \nabla S_u
\cdot \tau = 2\pi d N_v$ for vortices with winding number $d$. On
the other hand, if the vortex lattice mimics solid-body rotation
there is $\Gamma = 2\Omega A$ where $A$ is the area enclosed by the
contour $\mc{C}$. In this approximation, the vortex density per area
is $n_v = N_v/A = \Omega/(\pi d)$ and the area per vortex is $1/n_v
= \pi d /\Omega$ and so decreases with increasing $\Omega$. The
crucial ingredient in this argument is the assumption of a uniform
distribution of vortices. But this is justified only if the number
of vortices is very large, i.e. if $\Omega$ is very large which
means for the TF regime that, indeed, $N_v, \Gamma$, and $\Omega$
have to increase as $\vare \to 0$: Uniform distribution means that
$A$ is finitely large, i.e. bounded from below by a positive
constant (independent of $\vare$). Actually, $A$ is the whole
condensate domain and the contour $\mc{C}$ is the boundary of that
domain. So, from $\Gamma = 2\Omega A$ we see that $\Gamma \simeq
\Omega$, i.e. the circulation is of the same order than the angular
velocity if the vortices are
distributed uniformly. \\
On the other hand, considering the case that $N_v$ and $\Gamma$
respectively can be bounded from above by a finite constant
(independent of $\vare$), then vortices can not be distributed
uniformly but instead they form a polygonal lattice (see e.g. the
pictures in \cite{mad2}). So the above argument concerning
solid-body rotation gives only an upper bound for $\Gamma$. However,
this bound is still quite good in experimental realizations as is
demonstrated in \cite{dali}. In order to estimate roughly the order
of magnitude of $\Omega$ for a finite number of vortices to appear,
one may calculate the GP energy of a single vortex. This has been
done most often in the approximation of a homogeneous system (see
e.g. Refs. \cite{PS,pita}) or for a condensate in harmonic traps
(see e.g. Refs. \cite{lundh,aftdu}).
 In any case, the leading contribution comes from the
    angular kinetic energy. This can be already seen heuristically
    by
considering a vortex of circulation $\Gamma = 2\pi d$ and core
radius $\sigma \sim \vare$ which is located at the origin of a flat
trap with radius $R$. Writing the vortex in the form
      $v(r,\theta) = \varrho(r) e^{i\theta d}$
    with $\varrho(r) \sim r^d$ if $0 \leq r \leq \vare R$ and $\varrho(r) \sim R^{-1}$
      if $\vare R \leq r \leq R$,
 the kinetic energy is then
$\int |\nabla v|^2 \sim R^{-2}(d^2|\ln \vare| + C)$, whereas the
rotation term gives $-iv^* \mb{\Omega}(\vare)\cdot (\nabla v \times
\mb{r}) = -d\Omega(\vare)$. Thus one may expect a vortex of winding
number $d$ to appear when
 \bdi
 \int ( |\nabla v|^2 - iv^* \mb{\Omega}(\vare) \cdot (\nabla v \times \mb{r})) \sim
 \frac{d^2}{R^2}|\ln \vare| - d\Omega(\vare) < 0
  \mbox{ i.e. }
   \Omega(\vare) > C \frac{d}{R^2} |\ln \vare|
    \edi
and the constant $C$ is fixed by the external potential accordingly.
 Hence one vortex or a finite
   number of them are favourable to exist if the angular velocity is of the order $\Omega(\vare) \simeq
   C |\ln \vare|$. \footnote{For $|\ln \vare| \ll \Omega(\vare) \ll
1/\vare$ the number of vortices is no longer bounded as $\vare \to
0$ but the density is still not affected in leading order, see
\cite{CDY}.} Furthermore, from $\Gamma < 2\Omega
   A \leq C$ we get $A \leq C/\Omega$, that is
   the vortices are enclosed within a disc centered at the origin having a radius of the order
    \beq \label{abstand}
     r_v \leq \frac{C}{\sqrt{\Omega}} \simeq \frac{C'}{\sqrt{|\ln
     \vare|}}.
      \eeq
 So with regard to the above discussion, we model the vorticity in terms of a finite
number of vortices within the TF domain denoting their positions as
$\mb{r}_i=(x_i,y_i) \in \mc{D} \sm \p \mc{D},~ i=1,..,n$, $n \in
\mathbb{N}$. The vortex cores are modelled as non-overlapping discs
$B_i=B(\mb{r}_i,\sigma)$ with core radius $\sigma \sim \vare$,
 all contained within $\mc{D}$:
\beq \label{voc}
 \bar B(\mb{r}_i,\sigma) \subset \mc{D} \mbox{ for all
} i,~~\bar B(\mb{r}_i,\sigma) \cap \bar B(\mb{r}_j,\sigma) =
\emptyset \mbox{ for all } i \not= j \eeq
 assuming that $|\mb{r}_i-\mb{r}_j| > 2\sigma$.
 Otherwise, their energy would surpass the order of $|\ln \vare|$ and
would hence not be favourable for the angular velocities considered
here.
 In a vortex
point $\mb{r}_i$, the condensate wave function vanishes
$|u|(\mb{r}_i) = |v|(\mb{r}_i) = 0~\forall~i$ and the circulation
condition can be written as \beq \label{quant} \int_{\p B_i} \nabla
S_u \cdot \tau = \int_{\p B_i} \nabla S_v \cdot \mb{\tau} = \int_{\p
B_i} \frac{\p S_v}{\p \tau} = 2\pi d_i~~\forall~i, \eeq where
$\mb{\tau}$ is the unit tangent vector to $B_i$ and $d_i$ is the
degree of the vortex in $\mb{r}_i$.
 The domain outside the cores is denoted as
 \bdi
 \tilde{\mc{D}} := \mc{D}\setminus \bigcup_{i}  B_{i}.
 \edi
In that region, there holds $|u| \to f$, i.e. $|v| \to 1$ and we may
thus approximate
 \beq \label{vort1}
 0 \leq |v| \leq 1 - o(1) \mbox{ in } B(\mb{r}_i,\sigma)
 \eeq
 and
 \beq \label{vort2}
 |v| = 1 - o(1) \mbox { in } \tilde{\mc{D}}
 \eeq
where $o(1)$ goes to zero for $\sigma \to 0$ (i.e. $\vare \to 0$).
The detailed form of the error in $o(1)$ depends on the steepness of
the radial falloff of the vortex core profile. For the core radii we
are going to use, namely $\sigma = \vare^{\alpha}, \alpha
> 0$, the error due to the core profile is negligible within the
orders considered.

\section{Lower bound for the Ginzburg-Landau-type energy $\glf[v]$}

In this section, we consider the functional
  \beq \label{gv} \glf[\ve] =
 \int_{\mc{D}} \left[\frac{\fe^{2}}{2}|\nabla \ve|^{2} +
 \frac{\fe^{4}}{4\vare^{2}}(1-|\ve|^{2})^{2}\right]
 \eeq
 which is part of the energy splitting (\ref{tfsplitting}). Because of (\ref{pointwisa}), we can replace $\fe$
by $\sqrt{\tfm}$ in (\ref{gv}) and the error is of the order $o(1)$.
Using the polar decomposition $\ve = |\ve|e^{iS_{v_{\vare}}}$,
(\ref{gv}) is equivalent to \beq \label{gw} \glf[\ve] =
\int_{\mc{D}} \left[ \frac{\tfm}{2}[(\nabla |\ve|)^2 +
|\ve|^2(\nabla S_{v_{\vare}})^2] +
\frac{(\tfm)^2}{4\vare^2}(1-|\ve|^2)^2\right] - o(1). \eeq
Minimizing this at fixed $\tfm$ results in
 \bdi
  -\Delta |\ve| + |\ve|(\nabla S_{v_{\vare}})^2 -
  \frac{(\tfm)^2}{\vare^2}|\ve|(1-|\ve|^2) = 0
   \edi
 and
 \beq \label{vel}
  \nabla \cdot \left[|\ve|^2 \nabla S_{v_{\vare}} \right] = 0.
  \eeq
We have the following estimate:
\\
\\
\tb{Estimate 4:} Let $\fe$ be a minimizer of (\ref{vofree}), $\ue$ a
minimizer of (\ref{skalen}) and $\ve=\ue/\fe e^{iS}$. Let $\sigma =
C\vare^{\alpha}$ with constants $C,\alpha > 0$ and let $\ve$ satisfy
(\ref{voc}) - (\ref{vort2}) in presence of vortices in $\mb{r}_i$
having winding numbers $d_i$, $i=1,..,n$. Then, for $\vare$
sufficiently small and $\Omega \leq C|\ln \vare|$ asymptotically,
the GL-type energy can be bounded from below by
 \bdi
 \glf[\ve] \geq \pi
 |\ln \sigma|\sum_{i=1}^n d_i^2 \tfm(\mb{r}_i)  + \pi
 \ln \frac{\sigma}{\vare}  \sum_{i=1}^n
 |d_i|\tfm(\mb{r}_i)  -
 \edi
  \beq \label{glower}
   - \pi\sum_{i \not=j} d_{i}d_{j}\ln |\mb{r}_{i}-\mb{r}_{j}|\tfm(\mb{r}_i) +
   o(1).
 \eeq
\tx{Proof:}
\\
First we are going to estimate $\glf[\ve]$ in the vortex-free domain
where $|\ve| = 1-o(1)$. Then (\ref{gw}) reduces to \beq \label{smin}
\glf[\ve]|_{\tilde{\mc{D}}} =\frac{1}{2} \int_{\tilde{\mc{D}}} \tfm
(\nabla S_{v_{\vare}})^2 = \frac{1}{2} \int_{\tilde{\mc{D}}} \tfm
\mb{V}^2 \eeq
 up to an error of order $o(1)$
and we use the relation $\nabla S_{v_{\vare}} = \mb{V}$
   where $\mb{V}$ is the (linear) superfluid
velocity of the condensate (see (\ref{vel})). Indeed, it is shown in
Ref. \cite{lsy3} that BECs are 100 \% superfluid in their ground
state. Minimizing the functional with respect to $\mb{V}$ gives
 \beq
\label{euls} \tfm \nabla \cdot \mb{V} + \nabla \tfm \cdot \mb{V} = 0
 \mbox{ in } \tilde{\mc{D}}.
 \eeq
Because of the circulation condition (\ref{quant})
 \bdi
 2\pi d_i =
\int_{\p B_i} \nabla S_{v_{\vare}} \cdot \mb{\tau} = \int_{\p B_i}
\mb{V} \cdot \mb{\tau} = \int_{B_i} \nabla \times \mb{V} \cdot
d\mb{o}, \edi
 there is \beq  \label{strom2} \nabla \times \mb{V}(\mb{r}) = 2\pi
d_i \mb{\delta}(\mb{r}-\mb{r}_i) \mbox{ in } B_i
 \eeq
 denoting the oriented surface element as $d\mb{o}$.
 Since the cores $B_i$ are small and the TF density $\tfm$ is smooth in $\mc{D} \sm \p \mc{D}$,
 $\tfm$ is nearly constant within them, and we
 may approximate
 \beq \label{strom1}
 \nabla \cdot \mb{V} = 0 \mbox{ in } B_i.
 \eeq
This allows to define a stream function $\psi_{\vare}$, which is the
dual to the phase $S_{v_{\vare}}$, so $\nabla S_{v_{\vare}} = \nabla
\times \psi_{\vare}$. Using this form for $\mb{V}$ together with
(\ref{strom2}) and (\ref{strom1}), the stream function becomes
 \beq \label{phasev}
 \psi_{\vare}(\mb{r}) = -d_i \ln |\mb{r}-\mb{r}_i| \mbox{ in } B_i.
 \eeq
Now we calculate the integral
 \bdi \frac{1}{2}\int_{\tilde{\mc{D}}} \tfm \mb{V}^{2} = \frac{1}{2} \int_{\tilde{\mc{D}}}
 \tfm (\mb{V} \times \nabla \psi_{\vare}) \cdot d\mb{o} = \frac{1}{2} \int_{\tilde{\mc{D}}} \nabla
 \times[\tfm\psi_{\vare} \mb{V}]\cdot d\mb{o} -
 \edi
  \bdi
 -\frac{1}{2} \int_{\tilde{\mc{D}}} \psi_{\vare} \nabla \tfm \cdot \mb{V}
  -\frac{1}{2} \int_{\tilde{\mc{D}}} \psi_{\vare} \tfm (\nabla \cdot \mb{V}) \edi
 \bdi = \frac{1}{2}  \int_{\p \tilde{\mc{D}}} \tfm
 \psi_{\vare} \mb{V} \cdot \tau - \frac{1}{2} \int_{\tilde{\mc{D}}} \psi_{\vare} [\nabla
\tfm \cdot \mb{V} + \tfm (\nabla \cdot \mb{V})] \edi
 \bdi =
\frac{1}{2}  \int_{\p \mc{D}} \tfm \psi_{\vare} \mb{V} \cdot \tau +
\frac{1}{2}\sum_{i=1}^n
 \int_{\p B_i} \tfm \psi_{\vare} \mb{V} \cdot \tau
 \edi
 \beq \label{vorr}
 = \frac{1}{2}\sum_{i=1}^n \tfm(\mb{r}_i)
  \int_{\p B_i} \psi_{\vare} \mb{V} \cdot \tau + o(1),
\eeq where we used Stokes theorem, (\ref{euls}) and $\tfm = 0$ on
$\p \mc{D}$. Using again the fact that the cores $B_i$ are small and
$\tfm$ is smooth in $\mc{D} \sm \p \mc{D}$, we replace it by its
value in the core center $\tfm(\mb{r}_i)$ and the error is of the
order $o(1)$.
 \\
If we insert $\psi_{\vare}$ from (\ref{phasev}) we get \bdi
\frac{1}{2}\sum_i \tfm(\mb{r}_i) \int_{\p B_i} \psi_{\vare} \mb{V}
\cdot \tau = -\frac{1}{2}\sum_i \tfm(\mb{r}_i) d_i \int_{\p B_i} \ln
|\mb{r}-\mb{r}_i| \mb{V} \cdot \tau -
 \edi
 \beq \label{eh} - \frac{1}{2}\sum_i \tfm(\mb{r}_i) \sum_{j\not= i} d_j
\int_{\p B_i} \ln |\mb{r}-\mb{r}_j| \mb{V} \cdot \tau + o(1). \eeq
 The first term on the
r.h.s. describes the 'diagonal part' ($i=j$) and the second one the
'non-diagonal part' ($i\not=j$). Using $\ln |\mb{r}-\mb{r}_i| = \ln
\sigma$ and $|\mb{r}-\mb{r}_j| = |\mb{r}_i-\mb{r}_j| + o(1)$ for
$\mb{r} \in \p B_i$, there simply remains in each case the
circulation condition which gives $2\pi d_i$. So we have \beq
\label{zw} \frac{1}{2}\sum_i \tfm(\mb{r}_i) \int_{\p B_i}
\psi_{\vare} \mb{V} \cdot \tau = -\pi \ln \sigma \sum_i d_i^2
\tfm(\mb{r}_i) - \pi \sum_{i\not= j} d_i d_j \ln |\mb{r}_i -
\mb{r}_j|\tfm(\mb{r}_i). \eeq
 Since $\sigma \ll 1$, $-\ln \sigma$ can be replaced by
$|\ln \sigma|$ and we recover the first and third term in
(\ref{glower}). In this result, we see the familiar energy
dependence on the winding number squared $d^2$ and the logarithmic
divergence due to the vortex cores $|\ln \sigma|$ (see also the
approach in Ref. \cite{lundh} for the harmonic trap case). If there
are more vortices present than one, their interaction energy is
modelled by \beq \label{wint}
  W(\mb{r}_1,..,\mb{r}_n) = -\pi \sum_{i \not= j} d_i d_j \ln |\mb{r}_i-\mb{r}_j|\tfm(\mb{r}_i)
  \eeq
in (\ref{zw}). It has the form of a Coulombian interaction in
2-dimensional systems where vortices with
  the same sign of the winding number repel each other and vortices with opposite sign attract each
  other.
In Ref. \cite{BBH}, the analogue to this function is called
renormalized energy because it remains after the core energy, which
is the leading order,
 is separated:
  $W < |\ln \sigma|$ as long as $|\mb{r}_i-\mb{r}_j| > 2\sigma$.
  We also see that
  $W$ is bounded from below by a constant
 if all winding numbers have the same sign.
\\
\\
It remains to find a lower bound of $\glf[\ve]$ in the vortex cores
$B_i$: \beqn \label{gb} \lefteqn{\glf[\ve] |_{\bigcup_i B_i} =
\sum_i \int_{B_i} \left[\frac{\tfm}{2}|\nabla \ve|^2 +
\frac{(\tfm)^2}{4\vare^2}(1-|\ve|^2)^2\right] - o(1) {}}
\nonumber\\
& &{} \geq \sum_i \int_{B_i} \frac{\tfm}{2}|\nabla \ve|^2 - o(1) =
\sum_i \frac{\tfm(\mb{r}_i)}{2} \int_{B_i} |\nabla \ve|^2 - o(1)
 \eeqn
where we used again the fact that the TF density $\tfm$ varies only
of the order $o(1)$ in the small discs $B_i$. The integral over
$B_i$ can be estimated as follows: \bdi \int_{B_i} |\nabla \ve|^2
\geq \int_{B_i \sm B_{\vare}} |\nabla \ve|^2 \geq
\int_{\vare}^{\sigma}\int_0^{2\pi}\frac{1}{r^2}\left|\frac{\p v}{\p
\phi}\right|^2 r dr d\phi \edi with polar coordinates $(r,\phi)$ on
the annulus, $B_{\vare}$ a disc with radius $\vare$ centered at
$\mb{r}_i$, and we use the polar decomposition
 $v(r,\phi) = |v|(r)e^{id_i\phi}$ for a vortex with winding number $d_i$
 in the disc $B_i$ with radius $\sigma$. Using $\int_0^{2\pi}|\frac{\p v}{\p \phi}| \geq 2\pi |d_i|$
  and Cauchy-Schwartz inequality, we get
\beq \label{gb2}
\int_{\vare}^{\sigma}\int_0^{2\pi}\frac{1}{r}\left|\frac{\p v}{\p
\phi}\right|^2 drd \phi \geq 2\pi |d_i| \ln \frac{\sigma}{\vare}.
\eeq
 Combining (\ref{eh}), (\ref{zw}), (\ref{gb}) and (\ref{gb2}), we complete the
 proof of (\ref{glower}).

\section{The rotation energy $\rotf[v]$}

The estimate for the rotation term \beq \label{rfunc} \rotf[\ve] =
\int_{\mc{D}} i f^2 \ve^* \nabla \ve \cdot (\nabla S - \mb{\Omega}
\times \mb{r}) \eeq in (\ref{tfsplitting}) proceeds similar as in
\cite{serfaty}. Because of (\ref{pointwisa}), we replace $\fe$ by
$\sqrt{\tfm}$ in (\ref{rfunc}) and the error is of the order $o(1)$.
Then we get from (\ref{conta}) \beq \label{contin} \nabla \cdot
[\tfm (\nabla S - \mb{\Omega} \times \mb{r})] = 0 \mbox{ in }
\tilde{\mc{D}} \eeq from where we see that there is a real function
$\chi(x,y)$ satisfying \beq \label{chidef} \tfm (\nabla S -
\mb{\Omega} \times \mb{r}) = \Omega \nabla^{\perp}\chi \eeq with
$\nabla^{\perp}\chi = (-\p_y \chi, \p_x \chi)$, $\p_x = \frac{\p}{\p
x}$, ect. We impose the accompanying boundary condition $\chi = 0
\mbox{ on } \p \mc{D}$. To determine the auxiliary function $\chi$
we use (\ref{chidef}) and rewrite it as \bdi (\nabla S - \mb{\Omega}
\times \mb{r})^{\perp} = \frac{\Omega}{\tfm}\nabla \chi \edi
 where $\mb{r}^{\perp} = (-y,x)$ if $\mb{r} = (x,y)$.
 Applying the operator
$\nabla$, we get \bdi \p_x(\p_y S -\Omega x) + \p_y(-\p_x S - \Omega
y) = \Omega \nabla \cdot \left(\frac{\nabla \chi}{\tfm}\right) \edi
or
 \beq \label{chidiff} \nabla \cdot
\left( \frac{\nabla \chi}{\tfm}\right) = -2. \eeq
 We have the
following estimate for the rotation term:
\\
\\
\tb{Estimate 5:}
  Let $\fe$
be a minimizer of (\ref{vofree}), $\ue$ a minimizer of
(\ref{skalen}), $\ve = \ue/\fe e^{iS}$ and $\chi$ the solution of
(\ref{chidiff}). Let $\sigma = C\vare^{\alpha}$ with constants
$C,\alpha > 0$ and let $\ve$ satisfy (\ref{voc}) - (\ref{vort2}) in
presence of vortices in $\mb{r}_i$ having winding numbers $d_i$,
$i=1,..,n$. Then for $\vare$ sufficiently small and $\Omega \leq
C|\ln \vare|$ asymptotically, the rotation energy is
 \beq \label{rotest}
  \rotf[\ve] =  2 \pi \Omega
\sum_{i=1}^n d_i \chi(\mb{r}_i) + o(1). \eeq
 \tx{Proof:}
\\
  We can see that the contribution in the vortex cores becomes small:
\bdi
 \left|\sum_{i=1}^{n}
 \int_{B_i}i\tfm \ve^* \nabla \ve \cdot (\nabla S - \mb{\Omega} \times \mb{r}) \right|
 \leq \sum_{i=1}^{n} \int_{B_i}
 |\tfm| |\ve| |\nabla \ve| |\nabla S - \mb{\Omega} \times \mb{r}|
 \edi
 \bdi \leq  n \mu \Omega \left(\int_{B_i}|v_{\vare}|^2\right)^{1/2} \left(\int_{B_i}|\nabla v_{\vare}|^2\right)^{1/2}
 \leq C \sigma |\ln \vare|^{3/2} \leq o(1)
 \edi
where we used that $|\int_{B_i} |\nabla v_{\vare}|^2| \leq C |\ln \vare|$ which will be shown in (\ref{neben}).\\
For estimating the rotation term outside of the vortex discs, we use
again $|\ve| = 1-o(1)$ to get
 \beqn
 \label{vine}
  \lefteqn{\int_{\tilde{\mc{D}}} i \tfm \ve^* \nabla \ve \cdot(\nabla S - \mb{\Omega}\times \mb{r}) =
 \Omega \int_{\tilde{\mc{D}}} i \ve^* \nabla \ve \cdot \nabla^{\perp} \chi = -\Omega \int_{\tilde{\mc{D}}}  \nabla S_v \cdot \nabla^{\perp} \chi {}}
\nonumber\\
& &{} = -\Omega \int_{\tilde{\mc{D}}} \nabla^{\perp} \cdot (\chi
\nabla S_v) + \Omega\int_{\tilde{\mc{D}}} \chi \nabla^{\perp} \cdot
\nabla S_v = -\Omega \int_{\p \tilde{\mc{D}}} \chi \nabla S_v \cdot
\mb{\tau}
\nonumber\\
& &{} = -\Omega \int_{\p \mc{D}}\chi \frac{\p S_v}{\p \tau} +
\Omega\sum_{i=1}^n \int_{\p B_i}\chi \frac{\p S_v}{\p \tau}
\nonumber\\
& &{} = \Omega \sum_{i=1}^n \chi(\mb{r}_i)\int_{\p B_i} \frac{\p
S_v}{\p \tau} + o(1) = 2\pi \Omega \sum_{i=1}^n d_i \chi(\mb{r}_i) +
o(1) \eeqn where we used (\ref{chidef}), $\nabla^{\perp} \cdot
\nabla S_v = 0$, $\chi = 0$ on $\p \mc{D}$ and (\ref{quant}).
Furthermore, since the cores $B_i$ are small and $\chi$ is smooth in
$\mc{D} \sm \p \mc{D}$, we replace it by its value in the core
center $\chi(\mb{r}_i)$ and the error is of the order $o(1)$. We
thus arrive at (\ref{rotest}).
\\
\\
The expressions (\ref{glower}) and (\ref{rotest}) for the vortex
contributions suggest that the lowest vortex energy is attained for
vortices with winding number $d_i=1$ for all $i$,
 which will be explicitly shown in \tx{Section 8.2}. In estimating
 (\ref{glower}), we used $\nabla \times \mb{V} = 2\pi d_i \delta(\mb{r}-\mb{r}_i)$ in the
vortex core $B_i$,
 which is valid for any vortex core radius $\sigma$. On the other hand, due to the characteristic scale $\vare$ the core
 is not much larger than
  $\sigma \sim \vare^{\alpha}, \alpha > 0$. Then,
   the gradient term of $\glf[v]$ \tx{outside} the core dominates over the contribution of the core itself.
Concerning the interaction energy, one could first of all ask which
terms of $\glf[v] - \rotf[v]$ in the splitting of the functional
$\gpf[u]$ in (\ref{splitting}) will
 contribute to the interaction between vortices. We have just seen that the rotation energy outside of
 vortex cores is of the 'diagonal' form given in (\ref{vine}), whereas it is of order $o(1)$
 in the cores. So the interaction must be modelled by the GL-type energy $\glf[v]$. In addition, the interaction is
  only relevant in the domain outside the cores. There we have $|v|\simeq 1$ and only the
  gradient term of $\glf[v]$ plays the significant role.
      In particular, the form of the core and interaction energy in
      (\ref{glower}) was
deduced by minimizing (\ref{smin}) with respect to $\nabla
S_{v_{\vare}} = \mb{V}$. The core energy dominates as long as
$\sigma \ll |\mb{r}_i-\mb{r}_j|$ for all $i\not= j$, i.e. as long as
the vortex core size
 is much smaller than the distance between vortices. This is vastly fulfilled in the regime
 $\vare \to 0$ and $\Omega \simeq C |\ln \vare|$
 since then there is $\sigma \simeq C \vare^{\alpha},~ \alpha > 0$ whereas
 $|\mb{r}_i-\mb{r}_j| \geq C/\sqrt{|\ln \vare|}$ which will be shown in \tx{Section 8.3}.
\\
\\
\tb{Remark:} Our analysis may be compared with the works of
\cite{aftdu} and \cite{ignat1}. We start from the original energy
functional (\ref{energie}) and rescale it to arrive at
(\ref{engfunc}) and (\ref{skalen}) respectively. In
\cite{ignat1,ignat2}, a functional is used, motivated in
\cite{aftdu} and justified by the normalization condition, which
already in the beginning looks like a GL-type functional. The so
encountered additional term is 'thrown away' because it does not
contribute to the vortex energy. But one has to keep in mind that
the true leading order is $1/\vare^2$ which can be explicitly seen
in (\ref{skalen}). In the estimate of the lower bound of (\ref{gv}),
we consider equations (\ref{smin}) and (\ref{euls}). The behaviour
of $\mb{V}$ in the discs is derived from the quantization condition
(\ref{quant}) and is given in (\ref{phasev}) in terms of the stream
function. These ingredients are used in the estimate of
(\ref{vorr}). Instead, methods of Ref. \cite{BBH} are adapted in
\cite{ignat2} to the functional studied by considering a 'linear
problem' as in \cite{BBH}. By introducing a suitable function, it
gives eventually the interaction energy between vortices. Concerning
the forthcoming estimates, we will benefit from inequality
(\ref{chiest}). The corresponding equality for $s=2$ is used in
\cite{aftdu,ignat1}. In \cite{ignat1,ignat2}, it is assumed that
$\Omega \leq C |\ln \vare|$ asymptotically. Otherwise, there is no a
priori assumption on the fine structure of vorticity. We have argued
in \tx{Section 4} that the assumption of a non-zero circulation
which is bounded from above by a natural number independent of
$\vare$ leads to an angular velocity of the order $\Omega \sim |\ln
\vare|$. So, our assumptions on the vortex fine structure,
concerning number and size of vortex cores, are actually compatible
with this order of $\Omega$.

 \section{Upper bound for $\glf[v]-\rotf[v]$}

The energy without vortex is always larger or equal to $\gpf[\ue]$
and it can be used as a trial function for the whole energy (see
also the proof of \tx{Estimate 1}): \bdi \gpf[\ue] = \gpf[\fe
e^{iS}] + \glf[\ve] - \rotf[\ve] + o(1) \leq \gpf[\fe e^{iS}] + C +
o(1)
 \edi
so $\glf[\ve] - \rotf[\ve] \leq C + o(1)$. A more precise upper
bound is obtained as follows:
\\
\\
\tb{Estimate 6:} Let $\fe$ be a minimizer of (\ref{vofree}), $\ue$ a
minimizer of (\ref{skalen}) and $\ve = \ue/\fe e^{iS}$. Then, for
$\vare$ sufficiently small and $\Omega \leq C |\ln \vare|$
asymptotically, the vortex energy $\glf[\ve] - \rotf[\ve]$ can be
bounded from above by
 \beq \label{gupper}
  \glf[\ve] - \rotf[\ve] \leq \pi
|\ln \vare|\sum_{i=1}^k d_i \tfm(\mb{r}_i) - 2\pi \Omega
\sum_{i=1}^k d_i \chi(\mb{r}_i) + W(\mb{r}_1,..,\mb{r}_k) + C + o(1)
 \eeq
 where $d_i \geq 1$ for all $i$, $W(\mb{r}_1,..,\mb{r}_k)$ from (\ref{wint}) and $i,j=1,..,k$.
 \\
 \\
\tx{Proof:}\\
 We fix $k \geq 1, k \in \mathbb{N}$
vortex positions $\mb{r}_1,...,\mb{r}_k$ in $\mc{D}$, each is center
of a disc with fixed radius $R > 0$ but small such that the discs
are completely contained in $\mc{D}$ and do not overlap i.e. $\bar
B(\mb{r}_i,R) \subset \mc{D}$ and $\bar B(\mb{r}_i,R) \cap \bar
B(\mb{r}_j,R) = \emptyset$ for all $i \not= j$. We use the trial
function $\hat{v} = |\hat{v}|e^{i\hat{S}_v}$ with $|\hat{v}|=1$ in
$\tilde{\mc{D}}$ and
 \bdi \hat{v}(\mb{r}) =
\left\{\begin{array}{ll}
 \frac{\mb{r}-\mb{r}_i}{|\mb{r}-\mb{r}_i|} & \textrm{for $|\mb{r}-\mb{r}_i| \geq \vare$}\\
 \frac{\mb{r}-\mb{r}_i}{\vare} & \textrm{otherwise}
 \end{array} \right.
\edi in the discs $B_i(\mb{r}_i,R)$. Since $|\nabla \hat{v}|^2 =
(\nabla \hat{S}_v)^2$ in $\tilde{\mc{D}}$ and the phase is smooth
and bounded outside the discs with finite size, there is \bdi
\glf[\hat{v}]|_{\tilde{\mc{D}}} = \int_{\tilde{\mc{D}}}
\frac{\tfm}{2}(\nabla \hat{S}_v)^2 \leq C. \edi However, in the
discs there is
 \bdi \glf[\hat{v}]|_{\bigcup_i B_i} = \sum_{i=1}^k
\frac{\tfm(\mb{r}_i)}{2}\int_{B_i} |\nabla \hat{v}|^2 + \sum_{i=1}^k
\frac{(\tfm)^2(\mb{r}_i)}{4\vare^2} \int_{B_i} (1-|\hat{v}|^2)^2 +
o(1) \edi where \beqn \label{neben} \lefteqn{\int_{B_i} |\nabla
\hat{v}|^2 -4\pi = \int_{B_i\sm B_{\vare}} |\nabla \hat{v}|^2 =
\int_{B_i\sm B_{\vare}} \frac{1}{|\mb{r}-\mb{r}_i|^2} =
\int_0^{2\pi} \int_{\vare}^{R} \frac{r dr d\phi}{r^2-2r\vare \cos
\phi + \vare^2} {}}
\nonumber\\
& &{} = \pi \ln(r^2-\vare^2)|_{\vare}^{R} \leq \pi  \ln
r^2|_{\vare}^{R} = 2\pi |\ln \vare| + 2\pi \ln R, \eeqn
  and the other contributions are at most of the order of a constant. With the above trial function, the rotation energy in
$\tilde{\mc{D}}$ is the same as in (\ref{rotest}) apart from the sum
running from $i=1$ to $k$. The contribution inside the vortex discs
is simply \bdi
 \left|\sum_{i=1}^{k}
 \int_{B(\mb{r}_{i},R)}i\tfm  \hat{v}^* \nabla \hat{v} \cdot (\nabla S - \mb{\Omega} \times \mb{r}) \right|
 \leq \sum_{i=1}^{k} \int_{B(\mb{r}_{i},R)}
  |\tfm| |\hat{v}| |\nabla \hat{v}| |\nabla S - \mb{\Omega} \times \mb{r}|
 \edi
 \bdi \leq  k \mu \Omega \left(\int_{B_i}|\hat{v}|^2\right)^{1/2} \left(\int_{B_i}|\nabla \hat{v}|^2\right)^{1/2}
 \edi
 \bdi \leq k \mu \Omega (\pi (R^2-\vare^2) + 3\pi \vare^2/2)^{1/2}(2\pi |\ln \vare| +
 2\pi \ln R + 4\pi)^{1/2}
\leq C \vare |\ln \vare|^{3/2}.
 \edi
 Then, to recover (\ref{gupper}) we finally use the fact that $W(\mb{r}_1,..,\mb{r}_k)$ can be bounded from below by a constant.

\section{The anisotropic homogeneous trap}

In the above estimates (\ref{energynorothom}), (\ref{glower}),
(\ref{rotest}) and (\ref{gupper}), the external trap potential $V$
enters via the Thomas-Fermi density $\tfm$ which was not specified
until now. These estimates are valid as long as $V$ satisfies
(asymptotical) homogeneity (see also the remark at the end of this
section). As an application, we consider now the potential in
(\ref{potential}). The associated TF density is
 \beq \label{tfmm}
 \tfm(x,y) = \frac{1}{2}\left(\mu -
(x^2 + \lambda^2 y^2)^{s/2}\right).
 \eeq
 From (\ref{norm}) we have $\mu =
(\frac{s+2}{s}\frac{2\lambda}{\pi})^{s/(s+2)}$. For $s \to \infty$,
$\mu \to 2\lambda/\pi$, hence $\mu$ is always smaller than one. The
auxiliary function $\chi$ is determined from (\ref{chidiff}) to \beq
\label{homchi} \chi(x,y) = \frac{1}{1+\lambda^2} \left[
\frac{1}{s+2}(x^2+\lambda^2y^2)^{(s+2)/2} - \frac{\mu}{2}
(x^2+\lambda^2 y^2) + \frac{s}{2(s+2)}\mu^{(s+2)/s} \right]. \eeq It
can be estimated from above in terms of the TF density $\tfm$ by
\beq \label{chiest} \chi(x,y) \leq
\frac{1}{1+\lambda^2}\frac{s2^{2/s}}{s+2}(\tfm(x,y))^{(2+s)/s}, \eeq
where strict equality only holds for the harmonic trap $s=2$ ! This
upper bound will be very useful in \tx{Section 8.2} where the
winding numbers of vortices are derived. The phase $S$ can be
determined by inserting (\ref{homchi}) and (\ref{tfmm}) into
(\ref{chidef}) and is already given in (\ref{nosing}).

\subsection{The energy with one vortex}

The upper bound of the energy using (\ref{tfsplitting}) and
(\ref{gupper}) is  \bdi \gpf[\ue] \leq \gpf[\fe e^{iS}] + \pi|\ln
\vare|\sum_{i=1}^k d_i \tfm(\mb{r}_i) - 2\pi \Omega \sum_{i=1}^k d_i
\chi(\mb{r}_i) + C + o(1). \edi
 For a trial
function having one vortex with winding number $d = 1$ at the origin
we get \beq \label{ep} \gpf[\ue] \leq \gpf[\fe e^{iS}] +
\frac{\pi}{2} \mu |\ln \vare| - \frac{\pi s
\mu^{(s+2)/s}}{(1+\lambda^2)(s+2)}\Omega + C +o(1). \eeq
 The energy $\gpf[\ue]$ will be smaller than the
vortex-free energy $\gpf[\fe e^{iS}]$ if the r.h.s. of (\ref{ep}) is
smaller or equal to $\gpf[\fe e^{iS}]-o(1)$. Equivalently, the
angular velocity must fulfill
  \bdi
 \Omega \geq \Omega_1 + C + o(1)
 \edi
 where
  \beq
\label{krit} \Omega_1 = \frac{s+2}{s\mu^{2/s}}\frac{1+\lambda^2}{2}
|\ln \vare| =: C_1 |\ln \vare|. \eeq
  (Equation (\ref{krit}) has to be multiplied
by $(16\vare^4)^{1/(s+2)}$ in order to obtain the unscaled angular
velocity $\tilde{\Omega}_1$, see (\ref{skalom})). So for $\Omega
\geq \Omega_{1} + C + o(1)$, minimizers of $\gpf[u]$ will
 have vortices, or in other terms: $\Omega_1$ is the leading order in the angular velocity where the solution with one
vortex having $d=1$ starts to be globally thermodynamically stable.
We may also see the following: Consider $(x,y) \in \mc{D}$, let
$s>2$ and denote
 $\delta E = \pi |\ln
\vare|\tfm(x,y) - 2\pi \Omega \chi(x,y) + C$. Then one can see that
at the origin $\nabla (\delta E) (0,0) = \mb{0}$ and $\Delta (\delta
E)(0,0) = 4\pi\mu \Omega > 0$, i.e. $(0,0)$ is a local minimum for
\tx{all} $\Omega$. However, $(0,0)$ is a global minimum for $\delta
E (0,0) < 0$ or $\Omega_1 + C + o(1) < \Omega$ with $\Omega_1$ in
(\ref{krit}). However, for the harmonic trap $s=2$ one has $\nabla
(\delta E) (0,0) = \mb{0}$ and $\Delta (\delta E)(0,0) = -2\pi |\ln
\vare| (1+\lambda^2) + 4\pi \mu \Omega$. For $\Delta (\delta E)(0,0)
< 0$, the origin is a local maximum; for $\Delta (\delta E)(0,0)
> 0$ the origin is a local minimum. So there is an angular
velocity for \tx{local} thermodynamical stability which is $\Omega
> \frac{1+\lambda^2}{2\mu}|\ln \vare| = \Omega_1/2$ and $\Omega_1$
in (\ref{krit}) with $s=2$ which was also shown in Ref.
\cite{aftdu}.

\subsection{All vortices are single-quantized}

\tb{Estimate 7:} If $\sigma = \vare^{\alpha}, 0 < \alpha < 1$,
$\vare$ sufficiently small and $\Omega \leq \Omega_1 + C F(\vare)$
with $F(\vare)$ of lower order than $|\ln \vare|$, then $d_i = 1$
for all $i$.
\\
\\
\tx{Proof:} \\
 We use
$\sigma = \vare^{\alpha}$ with $0<\alpha<1$ in (\ref{glower}) and
(\ref{rotest}), $W \geq C$ and
 \bdi
 \Omega \leq C_1 |\ln \vare| + C_2 F(\vare)
 \edi
  with $C_1$ from (\ref{krit}) and $C_2$ is another positive
  constant.
This upper bound for $\Omega$ is suggested by (\ref{krit}) and
$F(\vare)$ is assumed to be of lower order than $|\ln \vare|$. In
the next section, we will see that $F(\vare) = \ln |\ln \vare|$. So
 \bdi \pi |\ln \sigma| \sum_{i=1}^n d_i^2 \tfm(\mb{r}_i) + \pi
\ln \frac{\sigma}{\vare} \sum_{i=1}^n d_i \tfm(\mb{r}_i) -2\pi
\Omega \sum_{i=1}^n d_i \chi(\mb{r}_i) \leq C \edi or \bdi \pi |\ln
\sigma| \sum_i (d_i^2-d_i) \tfm(\mb{r}_i) + \pi |\ln \vare| \sum_i
d_i \tfm(\mb{r}_i) \leq C + 2\pi \Omega \sum_i d_i \chi(\mb{r}_i)
\edi \bdi \leq C + \frac{2\pi
\Omega}{1+\lambda^2}\frac{s}{s+2}\mu^{2/s}\sum_i d_i \tfm(\mb{r}_i)
\edi \bdi \leq C + (C_1|\ln \vare| + C_2
F(\vare))\frac{2\pi}{1+\lambda^2}\frac{s}{s+2}\mu^{2/s}\sum_i d_i
\tfm(\mb{r}_i) \edi \beq \label{single} \leq C + \pi |\ln \vare|
\sum_i d_i \tfm(\mb{r}_i) + C' F(\vare) \sum_i d_i \tfm(\mb{r}_i).
\eeq Here we used (\ref{chiest}) and $(\tfm)^{(s+2)/s} \leq
\left(\frac{\mu}{2}\right)^{2/s}\tfm$. We also note that the last
inequality in (\ref{single}) is valid only for $C_1$ from
(\ref{krit}).
\\
So we see that (\ref{single}) reduces to \bdi \sum_{i=1}^n (d_i^2 -
d_i) \tfm(\mb{r}_i) \leq o(1) \edi
 for $\vare$ sufficiently small.
Therefore, if the vortices are not located at the boundary of the
Thomas-Fermi domain where $\tfm$ vanishes, there must be $d_i = 1$
for all $i$ for sufficiently small $\vare$.

\subsection{The energy with $n$ vortices}

Since all vortices have winding number one, we see from the lower
and upper
 bounds of $\gpf[\ue]$ in (\ref{glower}), (\ref{rotest}) and (\ref{gupper}) that they coincide
 in their orders (up to a constant).
By applying the transformation $\tilde
{\mb{r}}_i=(\tilde{x}_i,\tilde{y}_i)$ with $\tilde{x}_i = x_i
\sqrt{\Omega}, \tilde{y}_i = y_i \lambda \sqrt{\Omega}$, the vortex
interaction energy $W(\mb{r}_1,..,\mb{r}_n)$ in (\ref{wint}) can be
decomposed as follows:
 \beqn \label{renem}
 \lefteqn{W(\mb{r}_1,..,\mb{r}_n) = -\pi \sum_{i\not= j}\ln |\mb{r}_{i}-\mb{r}_{j}|\tfm(\mb{r}_i) {}}
 \nonumber\\
 & &{} = -\frac{\pi}{4} \sum_{i\not=
 j}\ln (|x_{i}-x_{j}|^{2}
 + |y_{i}-y_{j}|^{2})(\mu-(x_i^2+\lambda^2y_i^2)^{s/2})
 \nonumber\\
 & &{} = \frac{\pi}{4}\mu n(n-1) \ln \Omega -
 \frac{\pi}{4} \frac{\ln \Omega}{\Omega^{s/2}} \sum_{i} (\tilde{x}_i^2 + \tilde{y}_i^2)^{s/2} -
 \frac{\pi \mu}{4} \sum_{i \not= j} \ln \left((\tilde{x}_i-\tilde{x}_j)^{2} +
 \frac{1}{\lambda^{2}}(\tilde{y}_i-\tilde{y}_j)^{2}\right)
\nonumber\\
& &{} + \frac{\pi}{4\Omega^{s/2}} \sum_{i\not= j} \ln
\left((\tilde{x}_i-\tilde{x}_j)^{2} +
 \frac{1}{\lambda^{2}}(\tilde{y}_i-\tilde{y}_j)^{2}\right) (\tilde{x}_i^2 + \tilde{y}_i^2)^{s/2}
\nonumber\\
& &{} = \frac{\pi}{4}\mu n(n-1)\ln \Omega - \frac{\pi}{4}\mu
\sum_{i\not= j} \ln \left(|\tilde{x}_i-\tilde{x}_j|^{2} +
 \frac{1}{\lambda^{2}}|\tilde {y_{i}}-\tilde {y_{j}}|^{2}\right) + o(1).
 \eeqn
The first term contains the relevant order $\ln \Omega$ whereas the
remainder is of the order of a constant. The rotation term
(\ref{rotest}) becomes
 \beqn
 \lefteqn{-2\pi \Omega
 \sum_{i=1}^n \chi(\mb{r}_{i}) =   {}}
\nonumber\\
& &{} -\frac{2\pi \Omega}{1+\lambda^2}
 \sum_{i}\left[\frac{1}{s+2}(x_i^2+\lambda^2 y_i^2)^{(s+2)/2} - \frac{\mu}{2}(x_i^2 + \lambda^2 y_i^2) + \frac{s}{2(s+2)}\mu^{(s+2)/s} \right]
 \nonumber\\
 & &{} = -\frac{\pi s n}{(1+\lambda^2)(s+2)}\mu^{(s+2)/s}\Omega + \frac{\pi \Omega \mu}{1+\lambda^2} \sum_{i}(x_{i}^{2} +\lambda^{2}y_{i}^{2})
\nonumber\\
& &{} - \frac{2\pi \Omega}{(1+\lambda^2)(s+2)} \sum_i
 (x_i^2 + \lambda^2 y_i^2)^{(s+2)/2}
 \nonumber\\
 & &{} = -\frac{\pi s n}{(1+\lambda^2)(s+2)}\mu^{(s+2)/s}\Omega + \frac{\pi \mu}{1+\lambda^2}
 \sum_{i}(\tilde{x}_i^{2} + \tilde{y}_i^{2})
\nonumber\\
& &{} -
 \frac{2\pi}{(1+\lambda^2)(s+2)}\frac{1}{\Omega^{s/2}} \sum_i (\tilde{x}_i^2 + \tilde{y}_i^2)^{(s+2)/2}
 \eeqn
where in the last step the variable transformation
was applied.\\
 The remaining part of the lower bound of $\glf[\ve]$ in (\ref{glower}) is (with $d_i = 1~\forall i$)
  \bdi
\pi |\ln \vare^{\alpha}|\sum_i \tfm(\mb{r}_i) + \pi
 \ln (\frac{\vare^{\alpha}}{\vare})\sum_i \tfm(\mb{r}_i)
 = \pi |\ln \vare|\sum_i \tfm(\mb{r}_i)
 \edi
 for small $\vare$ and using $\sigma = \vare^{\alpha}$.
With $\ue$ and $\fe$ as above, we thus recover (\ref{nenergie}) for
the Gross-Pitaevskii energy in presence of $n$
 vortices
\bdi
 \gpf[\ue] = \gpf[\fe e^{iS}] + \frac{\pi}{2} \mu n\left(|\ln
\vare| - \frac{2s}{(1+\lambda^2)(s+2)}\mu^{2/s}\Omega\right)+
 \edi
 \beq \label{el}
+\frac{\pi}{4}\mu n(n-1) \ln \Omega +
w(\tilde{\mb{r}}_1,...,\tilde{\mb{r}}_n) + C + o(1) \eeq with
 \bdi
w(\tilde{\mb{r}}_1,...,\tilde{\mb{r}}_n) = - \frac{\pi
\mu}{4}\sum_{i\not=j}\ln\left ((\tilde{x}_i-\tilde{x}_j)^2 +
\frac{1}{\lambda^2}(\tilde{y}_i-\tilde{y}_j)^2\right) +
 \edi
\bdi + \frac{\pi \mu}{1+\lambda^2}\sum_i
(\tilde{x}_i^2+\tilde{y}_i^2) - \frac{\pi \ln
\Omega}{4\Omega^{s/2}}\sum_i (\tilde{x}_i^2+\tilde{y}_i^2)^{s/2}
\edi
 where we put all terms proportional to $ \Omega^{-m}, m
> 0$ into $o(1)$ since $\Omega \leq C|\ln
\vare|$ asymptotically.  \\
\\
 A necessary condition for the minimizing configuration to
have more than one vortex is \beq \label{gl}
 \min_{U_n} \gpf[u]  \leq \min_{U_1} \gpf[u]
\mbox{ for } n \geq 2 \eeq where $U_n$ is the set of functions with
$n$ vortices having winding number one each and $U_1$ is the set of
functions with one vortex at the origin with winding number one.
Using (\ref{gl}), we first want to deduce a rough estimate for the
critical angular velocity $\Omega_n$ for $n$ vortices to appear. To
this aim, we neglect in (\ref{el}) the term coming from the
interaction and we take the energy with all vortices close to the
origin, i.e. we approximate $\tfm(\mb{r}_i) \approx \mu/2$ for all
$i$, which is a more stringent condition on the l.h.s. of
(\ref{gl}). Indeed, from (\ref{abstand}) we expect the vortices to
be near to the origin. Hence \bdi \frac{\pi}{2} |\ln \vare|\mu (n-1)
+ \frac{\pi}{4}\mu n(n-1)\ln \Omega_n + \frac{\pi s
\mu^{(s+2)/s}}{(1+\lambda^2)(s+2)} \Omega_1 \leq \frac{\pi s n
\mu^{(s+2)/s}}{(1+\lambda^2)(s+2)} \Omega_n + C, \edi and \bdi
\frac{1+\lambda^2}{2}\frac{s+2}{s\mu^{2/s}}|\ln \vare| \frac{n-1}{n}
+ \frac{1+\lambda^2}{2}\frac{s+2}{s\mu^{2/s}}\frac{n-1}{2} \ln
\Omega_n  + \frac{1}{n}\Omega_1
 \leq
\Omega_n + C. \edi
 Using (\ref{krit}) and the fact that $\Omega_1 \leq \Omega_n$ for $n \geq 2$, we have the estimate
\bdi \Omega_1 +
\frac{1+\lambda^2}{2}\frac{s+2}{s\mu^{2/s}}\frac{n-1}{2}\ln \Omega_1
\leq \Omega_n + C \edi
 which can be put into the form
 \bdi
  \Omega_1 + C_1 \frac{n-1}{2}\ln |\ln \vare| + C_1 \frac{n-1}{2}\ln
  C_1 - C \leq \Omega_n
   \edi
 with $C_1$ and $\Omega_1$ from (\ref{krit}). We thus see that the
critical angular velocity has to be at least of the order $C_1|\ln
\vare| + C'\ln |\ln \vare|$ (we neglect the constant term). This
order of magnitude for $\Omega$ is assumed in Ref. \cite{ignat2}
from the outset before the number of vortices is rigorously derived.
Using now the ansatz \beq \label{rotrot2} \Omega = \Omega_1 +
C_1\nu(\vare) \ln |\ln \vare| \eeq with \bdi (k-1) + \delta \leq
\nu(\vare) \leq k - \delta \edi for an integer $k \geq 0$ counting
the number of vortices and $0 < \delta \ll 1$ a fixed constant
independent of $\vare$, we see the following: Inserting this form of
$\Omega$ in our energy estimate (\ref{el}) we get for the upper
bound  \bdi \gpf[\ue] \leq \gpf[\fe e^{iS}] + \frac{\pi}{2} \mu k
|\ln \vare| - \frac{\pi s k \mu^{1+2/s}}{(1+\lambda^2)(s+2)}\Omega +
\frac{\pi}{4}\mu k(k-1) \ln \Omega + C
 \edi
 \bdi
  = \gpf[\fe e^{iS}] - \frac{\pi}{2} k \mu \nu(\vare)\ln |\ln
\vare| + \frac{\pi}{4}\mu k(k-1)\ln |\ln \vare| + \frac{\pi}{4}\mu
k(k-1)\ln C_1 + C
 \edi
 or
 \bdi
 \glf[v_{\vare}] - \rotf[v_{\vare}] \leq - \frac{\pi}{2} \mu k \nu(\vare)\ln |\ln
\vare| + \frac{\pi}{4}\mu k(k-1)\ln |\ln \vare| +
 \edi
 \beq \label{alup}
+ \frac{\pi}{4}\mu k(k-1)\ln C_1 + C \eeq respectively.
 Considering the case $k=0$ (no vortices), i.e. $\nu(\vare)$ in
 (\ref{rotrot2}) satisfies $-1+\delta \leq \nu(\vare) \leq -\delta$,
 we see from (\ref{rotrot2}) and (\ref{krit}) that $\Omega \leq \Omega_1 - C_1\delta \ln |\ln
 \vare|$ and
  $\glf[\ve] - \rotf[\ve] = C$, showing (\ref{enga}).\\
 Considering $k=1$ (one vortex), i.e. $\delta \leq \nu(\vare) \leq 1-\delta$ and comparing its
  energy with the lower bound of
$\mc{G}_f[\ve]$ and $\mc{R}_f[\ve]$ we have \bdi - \frac{\pi}{2} \mu
\nu(\vare)\ln |\ln \vare| + C
  \geq \pi |\ln
\vare|\sum_{i=1}^n \tfm(\mb{r}_i) - 2\pi \Omega \sum_{i=1}^n
\chi(\mb{r}_i) + C
 \edi
\bdi
 \geq \pi \sum_{i=1}^n \tfm(\mb{r}_i)(-\nu(\vare)\ln
|\ln \vare| + C) \geq -\frac{\pi}{2} \mu n \nu(\vare)\ln |\ln \vare|
\edi
 so \bdi 1 - o(1) \leq n, \edi that is, there is at least one
vortex for
 \bdi
 \Omega_1 + C_1\delta \ln |\ln \vare| \leq \Omega \leq \Omega_1 + C_1(1-\delta)\ln |\ln \vare| = \Omega_2 - C_1\delta \ln
 |\ln \vare|
 \edi (using (\ref{herz})) and $\vare$ sufficiently small. Now we compare the lower and
upper bound of the energy if $k > 1$ is arbitrary large: the upper
bound is in (\ref{alup}), whereas the lower bound is
 \bdi
 \glf[v_{\vare}] - \rotf[v_{\vare}] \geq - \frac{\pi}{2} \mu n
\nu(\vare)\ln |\ln \vare| + \frac{\pi}{4}\mu n (n-1)\ln |\ln \vare|
+
 \edi
  \beq \label{allow}
   + \frac{\pi}{4}\mu n(n-1)\ln C_1 + C. \eeq
Comparison of (\ref{alup}) and (\ref{allow}) gives \bdi -n
\nu(\vare) + \frac{1}{2}n(n-1) + o(1) \leq -k \nu(\vare) +
\frac{1}{2}k(k-1) + o(1). \edi Assuming now that $n \leq k-1$, we
have \bdi \nu(\vare)(k-n) \leq \frac{1}{2}(k-n)(k+n-1) + o(1), \edi
so \bdi (k-1) + \delta \leq \nu(\vare) \leq \frac{1}{2}(k+n-1) +
o(1) \leq k - 1 + o(1) \edi
which is a contradiction for $\vare$ sufficiently small, since $\delta$ is a fixed constant.\\
On the other hand, assuming $n \geq k+1$ we have \bdi
\nu(\vare)(n-k) \geq \frac{1}{2}(n-k)(k+n-1) + o(1), \edi so \bdi k
- \delta \geq \nu(\vare) \geq \frac{1}{2}(k+n-1) + o(1) \geq k +
o(1) \edi which is again a contradiction for $\vare$ sufficiently
small. So we see that there are exactly $n \equiv k$ vortices for
$\vare$ sufficiently small and from this follows \bdi
\glf[v_{\vare}] - \rotf[v_{\vare}] = -\frac{\pi}{2} \mu n
\nu(\vare)\ln |\ln \vare| + \frac{\pi}{4}\mu n (n-1)\ln |\ln \vare|
+ \frac{\pi}{4}\mu n(n-1)\ln C_1 + C \edi
 and
 \bdi
 \Omega_n + C_1\delta \ln |\ln \vare| \leq \Omega \leq \Omega_{n+1}-C_1\delta \ln |\ln \vare|
 \edi
 by using (\ref{herz}). This completes the proof of the main result
 stated in the end of \tx{Section 2}.
\\
\\
\tx{Special cases:}
\\
\\
For the harmonic trap $s = 2$ there is
 \bdi \Omega_n =
\frac{1+\lambda^2}{\mu}[|\ln \vare| + (n-1)\ln |\ln \vare|].
 \edi
From (\ref{skalom}), the unscaled angular velocity is then \bdi
\tilde{\Omega}_n = \frac{2}{\mu}(1+\lambda^2)\vare [|\ln \vare| +
(n-1)\ln |\ln \vare|] \edi which may be compared to the results in
Refs. \cite{aftdu} and \cite{castin}. For the flat trap $s \to
\infty$, there is \bdi \Omega_n = \frac{1+\lambda^2}{2}[|\ln \vare|
+ (n-1)\ln |\ln \vare|] \edi and the same is true for
$\tilde{\Omega}_n$ since $\tilde{\Omega}_n \to \Omega_n$ for $s \to
\infty$. (We put in mind that for the flat trap, the scaled energy
functional converges to the original one, i.e.
$\mathcal{E}^{\mathrm{GP'}}[u'] \to \gpf[u]$, see \tx{Section 2}).
 By comparing the first critical angular velocities, we see the following: For the flat trap
 $\tilde{\Omega}_1
\sim |\ln \vare|$, resembling the corresponding velocity for the
rotating bucket and this is no surprise since the flat trap
approximates the bucket. On the other hand, there is
$\tilde{\Omega}_1 \sim \vare |\ln \vare|$ for the harmonic trap
which is much smaller. The ratio of the unscaled first critical
angular velocities is thus \bdi \frac{\tilde{\Omega}_1(s\to
\infty)}{\tilde{\Omega}_1(s=2)} = \frac{\mu}{4\vare}. \edi

\subsection{The vortex pattern}

The minimization of the energy in (\ref{el}) with respect to the
coordinates $\tilde{\mb{r}}_i = (\tilde{x}_i,\tilde{y}_i)$
determines the distribution of the vortices in the condensate and
therefore the resulting pattern which appears for a given number $n$
of vortices. The energy in (\ref{el}) is minimal with respect to the
coordinates if $w(\tilde{\mb{r}}_1,...,\tilde{\mb{r}}_n)$ is
minimal. Setting $\nabla w = \mb{0}$, we obtain
 \beq \label{pos1}
  \frac{\pi \mu}{2}
\sum_{i \not= j} \frac{\tilde{x}_i -
\tilde{x}_j}{(\tilde{x}_i-\tilde{x}_j)^2 +
\lambda^{-2}(\tilde{y}_i-\tilde{y}_j)^2} + \frac{\pi s \ln
\Omega}{2\Omega^{s/2}}\sum_i \tilde{x}_i(\tilde{x}_i^2 +
\tilde{y}_i^2)^{s/2-1} = \frac{2\pi \mu}{1+\lambda^2}\sum_i
\tilde{x}_i
 \eeq
 and
  \beq \label{pos2}
   \frac{\pi \mu}{2\lambda^2}\sum_{i \not= j}
\frac{\tilde{y}_i-\tilde{y}_j}{(\tilde{x}_i-\tilde{x}_j)^2 +
\lambda^{-2}(\tilde{y}_i-\tilde{y}_j)^2} + \frac{\pi s \ln
\Omega}{2\Omega^{s/2}}\sum_i \tilde{y}_i(\tilde{x}_i^2 +
\tilde{y}_i^2)^{s/2-1}= \frac{2\pi \mu}{1+\lambda^2}\sum_i
\tilde{y}_i.
 \eeq
 Multiplying (\ref{pos1}) and (\ref{pos2}) with
$\tilde{x}_i$ and $\tilde{y}_i$ respectively and adding them
together gives \beq \label{pos3} \sum_i (\tilde{x}_i^2 +
\tilde{y}_i^2) = \frac{1+\lambda^2}{4} \frac{n(n-1)}{2} +
\frac{1+\lambda^2}{4\mu}\frac{s \ln \Omega}{\Omega^{s/2}}\sum_i
(\tilde{x}_i^2+\tilde{y}_i^2)^{s/2}. \eeq On the other hand,
multiplying them with $\tilde{y}_i$ and $-\lambda^2\tilde{x}_i$
respectively and adding them gives \beq \label{pos4}
(1-\lambda^2)\sum_i \tilde{x}_i\tilde{y}_i =
\frac{1+\lambda^2}{4\mu}\frac{s \ln \Omega}{\Omega^{s/2}}
(1-\lambda^2)\sum_i \tilde{x}_i \tilde{y}_i
(\tilde{x}_i^2+\tilde{y}_i^2)^{s/2-1}. \eeq The relations
(\ref{pos3}) and (\ref{pos4}) are constraints for the
(non-dimensionalized) vortex positions.
 They simplify
considerably for the harmonic trap $s=2$ (and only for this trap!).
They were already deduced in Ref. \cite{aftdu} and we only state
them for completeness: $\sum_i \tilde{x}_i = \sum_i \tilde{y}_i = 0$
and \bdi \sum_i (\tilde{x}_i^2 + \tilde{y}_i^2) =
\frac{n(n-1)}{4\left(\frac{2}{1+\lambda^2} - \frac{\ln
\Omega}{\Omega \mu}\right)}, ~~\left(\frac{2}{1+\lambda^2} -
\frac{\ln \Omega}{\Omega \mu}\right)(1-\lambda^2) \sum_i \tilde{x}_i
\tilde{y}_i = 0. \edi For the anisotropic case $\lambda \not= 1$,
the last relation leads to $\sum_i \tilde{x}_i \tilde{y}_i = 0$. For
$n=2$ vortices, one already sees that $\tilde{x}_1 = -\tilde{x}_2$
and the same for the $\tilde{y}$-coordinates. Similarly one can
proceed for $n>2$ vortices (see Ref. \cite{aftdu} for a more
detailed discussion).
\\
However, for anharmonic traps with $s > 2$ the above relations are
more complicated but one may proceed in a similar way than for
harmonic traps. What can be seen immediately is the fact that, as
$\vare \to 0$, (\ref{pos3}) and (\ref{pos4}) reduce to \bdi
\sum_{i=1}^n (\tilde{x}_i^2 + \tilde{y}_i^2) =
\frac{1+\lambda^2}{4}\frac{n(n-1)}{2} + o(1) \edi and
 \bdi (1-\lambda^2)\sum_{i=1}^n \tilde{x}_i \tilde{y}_i = o(1) \edi
where $o(1) \sim \frac{\ln \Omega}{\Omega^{s/2}}$. Remarkably, the
distribution of vortices in anharmonic traps with $s > 2$ differs
only in this lower order from each other.
\\
\\
 \tb{Remark:}
  The analysis in
the foregoing sections holds generally for asymptotically
homogeneous traps according to Def.1.1 in Ref. \cite{lieb} which is
as follows: $V$ is asymptotically homogeneous of order $s>0$ if
there is a function $U$ with $U(\mb{r}) \not= 0$ for $\mb{r} \not=
0$ such that
 \bdi
 \frac{\gamma^{-s}V(\gamma \mb{r}) - U(\mb{r})}{1+|U(\mb{r})|} \to
 0~~~\mbox{ as } \gamma \to \infty
  \edi
 and the convergence is uniform in $\mb{r}$. $U$ is clearly uniquely
 determined and homogeneous of order $s$, i.e. $U(\gamma \mb{r}) =
 \gamma^s U(\mb{r})$ for all $\gamma \geq 0$.
\\
 In the case
that $V$ itself is homogeneous, there is $V \equiv U$. But if $V$
for instance is a harmonic-plus-quartic potential, $U$ contains only
the quartic contribution. Consider for example the following trap
\beq \label{anhom} V(x,y) = (x^2+\lambda^2 y^2)[1+\zeta
(x^2+\lambda^2 y^2)] \eeq
 with $\zeta \in (0,1)$ independent of $\vare$ describing the degree of anharmonicity.
 This trap
is \tx{asymptotically} homogeneous of order $s=4$. But since $V$ is
not homogeneous, equation (\ref{skalen}) is not exactly right.
However, using the above definition for asymptotically homogeneous
potentials, (\ref{skalen}) can be used if $V(\mb{r})$ is replaced by
$U(\mb{r}) + o(1)$. So only the leading contribution, i.e. the
asymptotically homogeneous one in the potential is 'visible'. In
order to see the anharmonic contribution of (\ref{anhom}) in our
regime, one would have to introduce an additional scaling parameter,
i.e. $\zeta$ would have to depend on $\vare$ (see for instance Ref.
\cite{afta}).

\section{Conclusions}

In this paper, we studied the Gross-Pitaevskii (GP) energy and
density for Bose-Einstein condensates confined in asymptotically
homogeneous traps which are subjected to an external rotation in the
Thomas-Fermi (TF) limit when the coupling parameter goes to
infinity. We derived by analytical estimates the leading order of
the GP energy and density, which are given by the corresponding TF
quantities, and the next orders due to vortices. In deriving the
contributions of the vortices, we estimated the relation between the
vortex core sizes and the considered magnitude of angular velocity.
As an example, we considered a very general anisotropic homogeneous
potential for which we calculated the critical angular velocities
for a finite number of vortices together with the associated GP
energy. We have shown that all vortices inside the Thomas-Fermi
domain are single-quantized and arranged in a polygonal lattice
whose shape can be deduced explicitly by a few simple constraints
satisfied by the vortex positions. In fact, the results may be used
to compare with experiments when the latter involve asymptotically
homogeneous traps in the TF regime. In this paper, we considered the
above trap for the reason of explicitness and because it
incorporates the harmonic and flat trap for which most experimental
results are available, but any trap potential satisfying
asymptotical homogeneity could be used.
\\
\\
\tb{Acknowledgments}
\\
The author thanks Jakob Yngvason and Michele Correggi for helpful
discussions. Financial support by the Austrian Science Fund FWF
under grant P17176-N02 is gratefully acknowledged.



\end{document}